\documentclass[12pt,onecolumn,draftcls]{IEEEtran} 

\usepackage{verbatim}
\usepackage{amssymb}
\usepackage{amscd}
\usepackage{amsmath}
\usepackage{graphicx}
\usepackage{epstopdf}
\usepackage{epsfig}
\usepackage{amsthm}

\newcommand{\RR}{\mathbb{R}}

\newcommand{\ba}{\mathbf{a}}
\newcommand{\bb}{\mathbf{b}}
\newcommand{\bc}{\mathbf{c}}

\newcommand{\bx}{\mathbf{x}}
\newcommand{\by}{\mathbf{y}}
\newcommand{\bz}{\mathbf{z}}

\newcommand{\bA}{\mathbf{A}}
\newcommand{\bB}{\mathbf{B}}

\newcommand{\bF}{\mathbf{F}}

\newcommand{\bH}{\mathbf{H}}

\newcommand{\cA}{\mathcal{A}}
\newcommand{\cB}{\mathcal{B}}
\newcommand{\cF}{\mathcal{F}}

\newcommand{\cP}{\mathcal{P}}
\newcommand{\cQ}{\mathcal{Q}}
\newcommand{\cU}{\mathcal{U}}
\newcommand{\cV}{\mathcal{V}}

\newtheorem{theorem}{Theorem}[section]
\newtheorem{corollary}{Corollary}[section]
\newtheorem{proposition}{Proposition}[section]
\newtheorem{lemma}{Lemma}[section]
\newtheorem{definition}{Definition}[section]
\newtheorem{example}{Example}[section]

\begin{document}
\title{Control Communication Complexity of Distributed Actions}
\author{Wing Shing Wong and John Baillieul
\thanks{W. S. Wong is with the Chinese University of Hong Kong, email: wswong@ie.cuhk.edu.hk.  John Baillieul is with the Boston University, email:
johnb@bu.edu.  Research work is partially supported by the Research Grants Council of the Hong Kong Special Administrative Region under project 417207 and 417105.  Also, JB acknowledges support from ODDR\&E MURI07 Program Grant Number FA9550-07-1-0528, and the National Science Foundation ITR Program Grant Number DMI-0330171; WSW acknowledges support from the National Natural Science Foundation of China under grant number 61174060.}}
\IEEEaftertitletext{\vspace{-2\baselineskip}} 
\maketitle

\begin{abstract}
Recent papers have treated {\em control communication complexity} in the context of information-based, multiple agent control systems including nonlinear systems of the type that have been studied  in connection with quantum information processing.  The present paper continues this line of investigation into a class of two-agent distributed control systems in which the agents cooperate in order to realize common goals that are determined via independent actions undertaken individually by the agents.  A
basic assumption is that the actions taken are unknown in advance to the other agent.
These goals can be conveniently summarized in the form of a {\em target matrix}, whose entries are computed by the control system responding to the choices of inputs made by the two agents.  We show how to realize such target matrices for a broad class of systems that possess an input-output mapping that is bilinear.   One can classify control-communication strategies, known as {\em control protocols}, according to the amount of information sharing occurring between the two agents.  Protocols that assume no information sharing on the inputs that each agent selects and protocols that allow sufficient information sharing for identifying the common goals  are the two extreme cases.  Control protocols will also be evaluated and compared in terms of cost functionals given by integrated quadratic functions of the control inputs.  The minimal control cost of the two classes of control protocols are analyzed and compared.  The difference in the control costs between the two classes  reflects an inherent trade-off between communication complexity and control cost.
\end{abstract}


\section{INTRODUCTION}\setcounter{equation}{0}

In \cite{Wong} and \cite{WB} the authors proposed the concept of {\normalsize\em control  communication complexity}
as a formal approach for studying a group of distributed agents exercising independent actions to achieve common goals.  For distributed cooperative systems, it is natural to expect that communication can help improve system performance, such as reducing the control cost.  In this paper, we demonstrate how the concept of control communication complexity can lead to an inherent estimate of the value of communication bits in reducing the control cost.  

Information-based control theory aims to deal with systems in which the interplay between control and communication are closely intertwined.
For some early work see for example \cite{De}, \cite{WBr1}, \cite{WBr2}, \cite{EM}, \cite{FZ}, \cite{LB}, \cite{NE}, and \cite{TM}.
In this paper, we investigate information-based systems controlled by two distributed agents.  In particular, we focus on cooperative control, the goal of which is for the agents, Alice and Bob,
to induce a system output that depends jointly on the controls they independently select from their respective finite sets of control inputs.  The problems treated below have extensions to settings in which the number of agents is larger than two, but such extensions are beyond the scope of the present paper.

The concept of multiple distributed selections of control actions from a specified set of possible choices has  not received much attention in the control literature until recent
work by the authors.  
While we believe this perspective
is novel, there are connections
with earlier work on cooperative decision making such as in the team decision problems treated in  \cite{YCHo} and \cite{AW}.  It also makes contact with (but differs from) a substantial body of work that has been devoted to extending the concepts of centralized control to the treatment of distributed and multi-agent systems.  Space does not permit a complete survey, but relevant work includes \cite{Yuksel} where the exchange of information is modeled and then used to reduce a certain non-classical stochastic optimization problem to a classical one.  The recent papers \cite{YB} and \cite{MS} address problems of distributed control with feedback loops closed through networks of communication-constrained data channels.

The aims and fundamental problems encountered in our work on control communication complexity are substantially different.  In what is reported below, the objective has been to extend ideas of communication complexity theory such that the cost of both communication and control is explicitly modeled and taken into account.  The models being proposed are abstractions of a variety of physical processes arising in diverse applications such as controlled quantum spin systems and motion control of robotic vehicles.  Our goal is to understand the general principles underlying communication by means of the dynamics of such systems.

While the distinction between our work below and previous work on decentralized control is real, the differences can be subtle as in the following example.  Suppose that two agents wish to find touring strategies in order to meet as quickly as possible,
as exemplified by the scenario of a mother and her child separated in a crowded park.   For that problem, a practical solution, not necessarily optimal, is to ask
the child to stay in one place and for the mother to conduct a complete tour of the park.  For a more complex situation,
consider Alice and Bob who jog in the same park at the same time every day. 
By tacit understanding, they wish their paths to cross or not to cross according to their moods
(the choices) as prescribed in the following table

\vspace{0.1cm}
\begin{center}
\small
\begin{tabular}{|l|l|l|}
\hline 
& Alice in good mood & Alice in bad mood \\
\hline
Bob in good mood & Paths cross& Paths do not cross \\
\hline
Bob in bad mood & Paths do not cross & Paths do not cross \\
\hline
\multicolumn{3}{c}{Table 1}  
\end{tabular}
\end{center}
If Alice and Bob can call each other to communicate their choice of inputs, the problem is trivial to solve.
However, if direct communication is not available or allowed, a basic question is whether it is possible for
Alice and Bob to follow different tour paths based on their moods to accomplish the stated
objective.  Moreover, if multiple feasible solutions exist, we are interested in identifying those that are optimal with regard to an appropriate metric, such as the total energy expended by the agents over a period of time encompassing many jogs in the park.

Although Table 1 bears resemblance to the payoff functions in classical game theory, one cannot over-emphasize the point that there is no optimization of the table values in our problem formulation.

A second example that is more closely aligned to the model studied here is the mobile sensor network positioning problem introduced in \cite{Wong}.
Consider a mobile sensor network, such as a network of remote sensing satellites.  The sensor
network serves two agents, Alice and Bob; each of them wants to monitor a geographic region
of interest selected from a pre-defined list.   The agents do not communicate to each other directly;
in fact, they may not even know the identity of the other party.   If for any given ordered pair of choices there is an optimal configuration for the sensors, it is natural to investigate whether one can design communication and control strategies, known as {\em control protocols}, for the agents to jointly maneuver the mobile sensors to the optimal configuration based on their own individual control selections.   Moreover, if such feasible control protocols do exist, a second objective is to find the optimal protocols under appropriately defined performance measures.

To fix ideas for subsequent discussions, there is assumed to be a control system with two input channels, one to be used by each of the two agents. 
The case of larger numbers of agents will be considered elsewhere.
Throughout the paper we index the finite collection of control inputs that Alice can send to the system by a set of labels $\cA \equiv \{1, \ldots, m \}$.  Similarly, a finite set $\cB \equiv \{1, \ldots, n \}$ is used to label the controls available to Bob.  It is assumed that  over many enactments of the protocol the inputs used by Alice and Bob will appear to be randomly chosen samples from uniformly distributed random variables with sample spaces $\cA$ and $\cB$.   If one represents the target output when Alice chooses $\alpha=i$ and Bob chooses  $\beta=j$ by $\bH_{ij}$, then the $m$-by-$n$ matrix
\begin{equation}
\bH=\left[ \bH_{ij} \right]
\end{equation}
provides a compact representation of the set of target outputs for all possible choices of inputs and will be referred to as the
{\em target matrix}.

While the structure of control protocols will be explained in the following section, a basic observation is that the input labels of the agents, $\alpha\in \cA$ and $\beta\in \cB$, are key inputs to these protocols.  Once $\alpha$ and $\beta$ are specified, following the basic premise of \cite{KN}, the control protocol is assumed to run to completion.
For a control
protocol, $\cP$, let $\bx(\cP(i,j),t)$ represent the state at time $t$ when $\alpha=i$ and $\beta=j$.
If the system output mapping is represented by $F$, the {\em feasibility problem} of {\em protocol-realizing control}
is to determine whether it is possible to design a control protocol, $\cP$, so that at the termination time, $T$,
the following condition is satisfied for all $i \in \{1, \ldots, m \}$ and $j \in \{1, \ldots, n \}$:
\begin{equation}
F(\bx(\cP(i,j),T))= \bH_{ij}.
\end{equation}

Solutions to the feasibility problem may involve control protocols that require observations of the
system state; even though there is no direct communication links between the agents, it is possible for
them to signal to one another via the dynamic system.   An example of such a control protocol was
discussed in \cite{WB}.   The number of bits exchanged during the execution of a control protocol
is a useful indicator of its complexity.   On the other hand, control protocols can also be evaluated in terms of the control cost incurred, for example as measured by the control energy required.   Although communication complexity and control cost seem unrelated, we will demonstrate that there is a close relation between the two.

To make our results concrete, we will focus on a class of systems whose input-output mappings are bilinear.   A prototypical example of such a system is the {\em Brockett-Heisenberg system}, which we hereafter refer to as the B-H system. 
The system is arises in  sub-Riemannian geometry  (\cite{rwb1}), and it shares essential features with models arising in nonholonomic mechanics (\cite{RB1},\cite{RB2}) and quantum mechanics (\cite{WB}).  For B-H systems one can characterize exactly when protocol-realizing control problems have feasible solutions in the absence of any communication between the agents.  Moreover, when a problem is feasible it is possible to determine the minimal control cost needed to achieve it.
On the other hand, in the case that the agents have partial information about each other's choice of inputs, it will sometimes be possible to decompose the control communication problem into simpler parts on each of which a smaller control cost can be calculated.  The concept of $\epsilon$-signaling will be introduced, by means of which the agents may share partial information with each other.  Pursuing this idea, it will be possible to estimate the minimal control costs for  two extreme scenarios, one without side communication or partial prior knowledge and the other with enough communication to allow the players to precisely compute intermediate results regarding the target matrix.  The two extreme scenarios suggest a natural framework for appraising the inherent value of a communication bit in terms of control cost.

The organization of the paper is as follows.  In section 2, we provide a description of the basic
model  as well as the definition of key concepts. 
In section 3, background results on a bilinear input-output system are presented.
In section 4 we describe how to transform the optimization of a single round protocol into a matrix optimization
problem, the solution to which is presented in section 5.  Implications of the result
for understanding  the trade-off issue between communication complexity and control cost is explained in
section 6.   Multi-round protocols are discussed in section 7. 
Section 8 provides a brief conclusion of the paper. 

\section{THE BASIC MODEL}\setcounter{equation}{0}

The dynamical systems considered here are inherently continuous time systems of the form:

\begin{equation}
\left\{
\begin{array}{lll}
\dfrac{d}{dt}\mathbf{x}(t)&=& \mathbf{d}(\mathbf{x}(t),u(t),v(t)),\;\mathbf{x}(0) = \mathbf{x}_0 \in \RR^N,
\vspace{0.1cm}\\
{\bf z}(t)&=&{\bf c}(\bx(t)) \in \mathbb{R},
\end{array}
\right.
\label{eq:basic0}
\end{equation}
where $\mathbf{d}$ is an arbitrary smooth vector field and  $u$ and $v$ are scalar control functions that, once chosen, are applied over a time interval of standard length $T$.   The output $z(\cdot)$ is sampled at discrete time instants $t_0<t_1<\dots$, where $t_0=0$ and $t_{k+1}=t_k+T$.  Information about the state is made available to the two agents through encoded observations $\bb_A(\bx(t))$ and $\bb_B(\bx(t))$ which are made at the same time instants.
The fact there are standard intervals of time over which observations and selected control inputs are applied allows the analysis  to make contact with prior work on information-based control of discrete-time systems.  Under our assumptions, we thus consider
the following simplified version of the model introduced in \cite{Wong}.
For $k = 1, 2, \ldots$, let $t_k$ represent the fixed time when observations are taken. Define
\begin{equation}
\bx_k = \bx(t_k).
\end{equation}
Observations of the state,  $\bb_A(\bx_k)$ and $\bb_B(\bx_k)$, are made available to each agent
as encoded messages, $Q_k^{(A)}$ and $Q_k^{(B)}$, consisting of finite bit-length codewords; in other words, the ranges of these quantization functions are finite sets.  
Computation and communication delays in reporting observations to the agents are assumed to be negligible.
The agents select respective control actions,
$u_k = P_k^{A}(\cQ_k^{(A)}, \alpha)$ and $v_k = P_k^{B}(\cQ_k^{(B)}, \beta)$,
where $\cQ_k^{A}$ represents the set of coded observations
$\{Q_k(\bb_A(\bx_k)), \ldots, Q_1(\bb_A(\bx_1)) \}$; $\cQ_k^{B}$ is defined similarly.  Note that the arguments of $u_k$ and $v_k$ are defined on finite
discrete sets.  It is assumed that for each element in the corresponding argument set, $u_k$ identifies a unique element in a set of admissible Lebesgue measurable control functions over $[t_k, t_{k+1})$; similar assumption holds for $v_k$.
If one imposes a probability measure on the sets of choices, these controls may be interpreted in terms of the observation sigma algebras over the corresponding probability space.  However, this probabilistic interpretation is not essential to the results presented in this paper.
 
For reasons of simplicity, we assume that the controls $u$ and $v$ are scalar functions.  The codewords identifying the selected controls are transmitted to the dynamic system.
Computation and communication delays associated with this step are also assumed to be negligible.
Thus at time $t_k$ the dynamic system can determine the control selected by Alice and Bob.

The state transition (\ref{eq:basic0}) between times $t_k$ and $t_{k+1}$ is thus described by the following discrete time control model
where the controls are square integrable scalar functions:
\begin{equation}
\left\{ \begin{array}{l}
\bx_{k+1} = \ba(\bx_{k},  u_{k}, v_{k}), \quad \bx(0) = \bx_0 \in \RR^N,\\
\by_k^{(A)}=\bb_A(\bx_k) \in \RR^{\ell_A}, \;
\by_k^{(B)}=\bb_B(\bx_k) \in \RR^{\ell_B},\\
u_k = P_k^{A}(\cQ_k^{(A)}, \alpha), \;
v_k = P_k^{B}(\cQ_k^{(B)}, \beta),\\
\bz_k=\bc(\bx_k) \in \RR.
\end{array}
\right.
\label{eq:basic1}
\end{equation}

The quantity $\bz_k=\bc(\bx_k)$ is a global system output that is observable to Alice, Bob, and possibly to
exogenous observers as well.  The protocol parameters, $\alpha$ and $\beta$, are specified
at time $t_0$ and remain unchanged while the protocol runs to completion.  The case where these parameters are
allowed to change over time is an interesting extension which is not considered here.

\begin{definition}
A control protocol, $\cP$, consists of the functions:

$\{Q_k^{(A)}\}_{k=0}^\infty, \{Q_k^{(B)}\}_{k=0}^\infty, \{P_k^{(A)}\}_{k=0}^\infty$, and $\{P_k^{(B)}\}_{k=0}^\infty$.
\end{definition}

We define the epoch between time $t_{k}$ and $t_{k+1}$ as round $k+1$.
In round $k+1$, an agent first observes and then selects a control to be applied to the
system.  The selected controls $u_k=P_k^{A}(Q_k^{(A)}(\bb_A(\bx_k)), \alpha)$ and 
$v_k = P_k^{B}(Q_k^{(B)}(\bb_B(\bx_k)), \beta)$ will typically be time-varying in the epoch between time $t_k$ and $t_{k+1}$, but they depend only on the encoded value of the state at time $t_k$.  Hence, the agents use essentially open-loop controls during the round, but the selection of
the control at the beginning of the round can depend on partial state information.  

\begin{definition}
Consider a dynamic system, (\ref{eq:basic1}), with parameters defined  by $(\ba, \bb_A, \bb_B, \bc, \bx_0)$.
A target matrix $\bH$ is said to be realizable at termination time $T_f$
if there exists a $k$-round protocol, such that
\[ t_{k} = T_f, \]
and for any choice of indices $i \in \cA$ and $j \in \cB$, the $k$-round protocol defined with
$\alpha = i$ and $\beta=j$  terminates with
\begin{equation}
\bc(\bx_{k})=\bH_{ij}.
\end{equation}
\end{definition} 

The initial state is assumed to be fixed and known to the agents.
For the first round, there is no need for any communication to the agents.  That is, we assume the
quantization functions $Q_0^{(A)}$ and $Q_0^{(B)}$ always take the same value for all control choices
and need not be transmitted.

Because a target matrix may be realized by different protocols, there is an interest in identifying those that are optimal
with respect to some performance measures.  One such measure
is to count the number of communication bits exchanged during the protocol execution.
To make this precise, \cite{Wong} introduced the concept of  {\it control communication complexity} by extending the concept of  {\em communication complexity} that was introduced in computer science by Yao
\cite{yao1}.  Briefly speaking, given a dynamic system and a target matrix, one defines the protocol complexity of a feasible protocol to be the maximum number of bits exchanged by the agents in running the protocol to completion.   The control communication complexity is then the minimum protocol complexity over the set of all feasible protocols.  A caveat: unlike classical communication complexity, control communication complexity is defined with regard to a fixed dynamical system. 

In the models considered in the present paper, control inputs are square integrable functions, and this suggests measuring the complexity of a protocol in terms of the integral of a quadratic function of the control.   This makes contact with performance measures commonly used in centralized control.  Indeed, control communication complexity provides a rich new class of optimal control problems. 
While the relation between communication complexity and integral control cost may not be transparent,
the analysis of the B-H system in \cite{WB} demonstrates that there is an intriguing relation between them.   
Intuition suggests that protocols in which there is limited communication between the agents {\em except for their common observations of the system dynamics} may require a larger integral control cost than those that employ
a large number of communication bits in addition to the system observations.  If the agents have partial information about each others choices of inputs, control laws can be more precisely tailored.
A contribution of this paper is to take the first step towards analyzing this trade-off
by comparing two limiting types of control protocols, namely, single round protocols
which entail no communication bits, and protocols in which agents share partial information about which elements $\alpha\in\cA$ and $\beta\in\cB$ are governing the execution of the protocol.  We provide a fairly complete treatment of single round protocols, but multi-round protocols are less well understood at this point.   As a lower bound on the control cost, we assume that it is possible for the agents to communicate their choices  of inputs to each other without incurring any control cost or state change,
thereby decomposing the original problem into a sequence of standard optimal control problems.   The difference in the minimal control
costs between this type of protocol and the single round protocols having no communication provides an upper bound on the saving afforded by the bits of information that are communicated between the two agents as described in \cite{Wong}. 

In the current model the only unknown parameter to Alice is Bob's choice of control input and vice versa
the only unknown parameter to Bob is Alice's choice of input.  The single choice problem in which each agent has only one choice of control input is equivalent to the scenario where the agents have complete information on
the system.   Since the initial state is known and there is no state noise, optimal distributed control can be devised without
resorting to any state observations.  In other words, for optimizing single choice  problems,
we only need to consider single round protocols.  

For the rest of the paper we focus on scalar output functions to simplify the analysis.
Vector-valued outputs, as necessitated by the sensor network example, present a significantly
larger technical challenge, and will not be discussed here.

To analyze the cost of letting our control system evolve under different input curves in single round protocols,
we lift the evolution dynamics (\ref{eq:basic1}) back up to the continuous domain as described by equation (\ref{eq:basic0}) but with $\bc$ replaced
by a scalar function.

Unless stated otherwise, the controls $u$ and $v$ exercised by the agents are assumed to lie in a closed subspace $\mathcal{L}\subset L^2[0,T]$.
Let $\mathcal{L} \otimes\mathcal{L}$ represent the tensor product Hilbert space with inner product defined by
\begin{equation}
<u_1 \otimes v_1, u_2 \otimes v_2>=<u_1, u_2><v_1, v_2>.
\end{equation}
At time $T$, the input-output mapping of system (\ref{eq:basic0}) can be regarded as a functional from 
$\mathcal{L}\otimes\mathcal{L}$ to $\mathbb{R}$, denoted by $F$. $F$ of course depends
on the initial state $\bx_0$ but as the state is assumed to be fixed and since for the time being we consider
only single round protocols, this dependency can be hidden for simplification.
Without loss of generality we assume that $t_0=0$ and $t_1=T=1$.  

$F$ is a bounded functional if there exists a finite $\|F\|$ so that for all $(u,v) \in \mathcal{L}\otimes\mathcal{L}$,
\begin{equation}
|F(u,v)|  \leq \|F\| \|(u \otimes v)\|_{\mathcal{L}\otimes\mathcal{L}} = \|F\| \|u\|_{\mathcal{L}} \|v\|_{\mathcal{L}}.
\label{eq:boundedF}
\end{equation}

To realize a given target matrix $\bH$, the optimal controls in general depend jointly on the 
parameters, $\alpha$ and $\beta$.  However, if the agents make their choices independently and there is no communication
between them, $u$ can only depend on Alice's parameter $\alpha$ and $v$ on Bob's parameter $\beta$.

A target matrix $\bH$ is realized by a single round protocol $\cP$ if there exist sets of 
controls, $\cU=\{ u_1,\ldots, u_m \}$ and $\cV=\{ v_1,\ldots, v_n \},$ so that
\begin{equation}
F(u_i,v_j)=H_{ij}.
\end{equation}
We emphasize that $F$ represents the system output at time $T$.
Such a single round protocol solution may not always exist as we will see in subsequent sections.
The problem may become feasible if Alice and Bob can exchange information about their choices to each other.

Given the system (\ref{eq:basic0}) and the parameter sets $\cA$ and $\cB$, the cost of a single round control protocol $\mathcal{P}$ is
defined as an average of the required control energy, given explicitly by the formula,
\begin{equation}
I(\mathcal{U},\mathcal{V})=
\dfrac{1}{m}\sum_{i=1}^m \int_0^1 u_i^2(t)dt + \dfrac{1}{n}\sum_{j=1}^n \int_0^1 v_j^2(t)dt.
\label{eq:control cost}
\end{equation}
One can also write equation (\ref{eq:control cost}) in the form,
\begin{equation}
I(\mathcal{U},\mathcal{V})=
\dfrac{1}{mn}\sum_{i=1}^m \sum_{j=1}^n \left( \int_0^1 u_i^2(t)dt + \int_0^1 v_j^2(t)dt \right),
\label{eq:control cost2}
\end{equation}
which highlights the fact that the control cost is averaged over all possible event outcomes based on the control actions that are chosen by the agents.

In subsequent sections, we compute the minimum averaged control energy for an arbitrary target function $\mathbf{H}$.
That is, our aim is to compute
\begin{equation}
\hat{C}_F(\mathbf{H}) \equiv \min_{\mathcal{U}, \mathcal{V} \subset \mathcal{L}}  I(\mathcal{U},\mathcal{V})
\end{equation}
subject to the constraints that for $i=1, \ldots, m$ and $j=1, \ldots, n$,
\begin{equation}
F(u_i, v_j) = H_{ij}.
\end{equation}

Before concluding this section we note that there are some similarities between the cooperative control communication protocols studied in this paper and more classical dynamic game strategies as studied in, say,  \cite{BO}.  Yet there are fundamental differences; for example, optimization of payoff functions is not the focus of our investigation.
Using the rendezvous problem as an example, once the moods of the the agents are fixed, the outcome to be achieved is
automatically defined by the target matrix.  The investigation focus is on how to ensure that the target objective---paths crossing or not crossing---can be guaranteed.  The work here also has some connection with the many papers in the literature
dealing with distributed control of mobile agents, multi-agent  consensus problems, and classical team decision theory.  See,  for instance,
 \cite{CB},\cite{CMB},\cite{DM},\cite{FM},\cite{YCHo}, and \cite{OM}.  In these papers the dynamics of the subsystems controlled by the agents are usually not tightly coupled
and the control cost is not explicitly calculated, unlike the models we are considering here.  Moreover, allowing agents to select controls from sets of standard inputs is also a fundamental point of departure.

\section{SYSTEMS WITH BILINEAR INPUT-OUTPUT MAPPINGS}\setcounter{equation}{0}
The simplest system defined in Section 2 is probably of the following linear type:
\begin{equation}
\left\{
\begin{array}{l}
\dfrac{d\mathbf{x}(t)}{dt} = \mathbf{Ax}(t) + u(t)\mathbf{b}_1
+ v(t)\mathbf{b}_2,\:\mathbf{x}_0 \in \RR^N,
\vspace{0.05cm}\\
y_A(t) = y_B(t) = z(t) = \mathbf{c}^T \mathbf{x}(t) \in \mathbb{R}.
\end{array}
\right.
\end{equation}
The input-output mapping of this system is affine in $(u, v)$.
Moreover, it is easy to check that
if $F(u_i, v_k) = H_{ik}$,
$F(u_i, v_l) = H_{il}$, $F(u_j, v_k) = H_{jk}$, and
$F(u_j, v_l) = H_{jl}$ for some $u_i$, $u_j$, $v_k$, and $v_l$, then
\begin{equation}
H_{ik} - H_{jk} = H_{il} - H_{jl}.
\end{equation}
Thus, there are severe restrictions on the set of realizable target functions for such an input-output mapping.  
A class of distributed control systems that realizes a richer class of input-output mappings has two independent input channels entering the system in a jointly bilinear fashion.  More precisely, we have the following definition:
\begin{definition}
Consider a system defined by equation (\ref{eq:basic0}) with control functions
defined in $\mathcal{L}$, a closed subset of $L^2[0,T]$.  The system
is a {\em bilinear input-output system} if for any time $t \geq t_0$, the output at $t$, regarded as a mapping
$(u(\cdot ),v(\cdot ))\mapsto z(T)$ from
$\mathcal{L}\otimes\mathcal{L}$ to $\mathbb{R}$ is bilinear in the control function ordered pair $(u,v)$.
\end{definition}

In state-space representation, two prototypical classes come to mind.  The first class is:

\begin{equation}
\left\{
\begin{array}{l}
\dfrac{d\mathbf{x}(t)}{dt} = \mathbf{Ax}(t) + u(t)v(t)\mathbf{b},\:\mathbf{x}(0)=\mathbf{0} \in \RR^N,
\vspace{0.05cm}\\
y_A(t) = y_B(t) = z(t) = \mathbf{c}^T \mathbf{x}(t) \in \mathbb{R},
\end{array}
\right.
\label{eq:wong:GeneralBiLinear}
\end{equation}
for some matrix $\bA$ and vectors $\bb$ and $\bc$.

A second, in a way more important class, comes from the Brockett-Heisenberg (B-H) system and their generalization, \cite{rwb1},\cite{rwb2}.  The B-H system, denoted by ${\mathbf{\Sigma}}_B$,
can be described as follows. 
\begin{equation}
\left\{
\begin{array}{l}
\dfrac{d}{dt}\left( 
\begin{array}{l}
x\\y\\z
\end{array}
\right) = \left( 
\begin{array}{c}
u\\v\\vx-uy
\end{array}
\right) , \left( 
\begin{array}{l}
x(0)\\y(0)\\z(0)
\end{array}
\right)  =
\left( 
\begin{array}{l}
0\\0\\0
\end{array}
\right)  \in  \,\mathbb{R}^3,  \vspace{.1cm} \\
y_A(t)=y_B(t)=z(t) \equiv c((x(t),y(t),z(t)).
\end{array}
\right.
\label{eq:wong:BI}
\end{equation}

The fact that the input-output mappings of these systems are bilinear in $u$ and $v$ can be easily verified.  Note that in both cases we could take
$\mathcal{L}$ to be $L^2[0,T]$ which results in a bilinear
input-output function with an infinite dimensional matrix representation.
On the other hand, we could also restrict
$\mathcal{L}$ to some finite dimensional subspace of $L^2[0,T]$ if
only certain control functions are allowed to be used in the system.



While our main results make use 
of properties that are to some extent particular to (\ref{eq:wong:BI}), it is important to note that the system (\ref{eq:wong:BI}) has features of
an intrinsically geometric nature in common with a much larger class of two-input ``drift-free'' control systems whose output mappings are not necessarily bilinear functionals of the inputs.  These systems arise in applications including the kinematic control of nonholonomic wheeled vehicles (See, e.g.\ \cite{LeonardKrishna}.) and the control of ensembles of spin systems arising in coherent spectroscopy and quantum information processing. (See \cite{LiKhan2}.) 
To illustrate how important features of a broad class of two-input systems are held in common with (\ref{eq:wong:BI}), we consider the Bloch equations in a spatially fixed frame in which there is an rf-inhomogeneity but no Larmor dispersion.  The system is governed by ordinary
differential equations evolving on the rotation group $SO(3)$.
\begin{equation}
\dot X = [u(t)\Omega_y + v(t)\Omega_x] X, \ \ X(0)=I,
\label{eq:jb:BlochEq}
\end{equation}
where
\[
\Omega_x=\left(
\small
\begin{array}{ccc}
0&0&0\\
0&0&-1\\
0&1&0\end{array}\right),\ \ 
\Omega_y=\left(
\small
\begin{array}{ccc}
0&0&1\\
0&0&0\\
-1&0&0\end{array}\right),\]\\
\[ \ {\rm and}\ \
\Omega_z=\left(
\small
\begin{array}{ccc}
0&-1&0\\
1&0&0\\
0&0&0\end{array}\right)
\]
are the infinitesimal generators of rotations about the $x$-,$y$-, and $z$-axes in $\mathbb{R}^3$.  
To briefly explore the connection between (\ref{eq:wong:BI}) and (\ref{eq:jb:BlochEq}), we recall the {\em area rule}.

\medskip
\begin{lemma} {\rm (Brockett, \cite{rwb3}.)} Let $u,v\in L^2[SO(2)]$ be as above, and let $x,y$ be defined by (\ref{eq:wong:BI}).  Then $x,y\in L^2[SO(2)]$, and $z(\cdot)$ as defined by (\ref{eq:wong:BI}) satisfies 
\[
z(1) = \int_0^1 v(t)x(t)-u(t)y(t)\,dt = \oint x\,dy-y\,dx = 2A,
\]
where $A$ is the signed area enclosed by $(x(t),y(t))$.
\label{thm:jb:Area}
\end{lemma}

The connection between (\ref{eq:wong:BI}) and (\ref{eq:jb:BlochEq}) is made by the following.
\medskip
\begin{example}\rm
Let the inputs
\[
u(t)=\epsilon\cos(2\pi t ), \ \ \ v(t) = \epsilon\sin(2\pi t)
\]
be applied to both (\ref{eq:wong:BI}) and (\ref{eq:jb:BlochEq}).  For (\ref{eq:wong:BI}), corresponding to these inputs,
\[
x(t)=\frac{\epsilon}{2\pi}\sin(2\pi t)\ \ \ {\rm and}\ \ \ y(t)=-\frac{\epsilon}{2\pi}\cos(2\pi t).
\]
By Lemma \ref{thm:jb:Area} (or a direct calculation), the output of (\ref{eq:wong:BI}) is $z(1)=\epsilon^2/(2\pi)$.  For this choice of input, (\ref{eq:jb:BlochEq}) may be solved explicitly yielding
\[
\small
X(t)=\exp\left[\left(\begin{array}{ccc}
0&-2\pi &0\\
2\pi & 0 & 0\\
0&0&0\end{array}
\right)t\right]\exp\left[\left(\begin{array}{ccc}
0&2\pi &\epsilon\\
-2\pi & 0 & 0\\
-\epsilon&0&0\end{array}
\right)t\right].
\]
Note that the first factor in this expression has period $1$, while the second factor is periodic with period $T=1/\sqrt{1+(\epsilon/(2\pi))^2}$.  At time $T$,
\[
\small
X(T)=\left(\begin{array}{ccc}
\cos\theta & \sin\theta & 0\\
-\sin\theta & \cos\theta & 0\\
0 & 0 & 1\end{array}
\right)
\]
where $\theta=-(1/2)(\epsilon^2/(2\pi))+0(\epsilon^2)$.  Hence, the output of (\ref{eq:wong:BI}) provides a good approximation of the $z$-axis rotation $\theta=-z(1)/2$ resulting from application the inputs $u,v$ to (\ref{eq:jb:BlochEq}) over the time interval $[0,1/\sqrt{1+(\epsilon/(2\pi))^2}]$.
We refer to \cite{WB} for further information regarding area rules and approximate motions of (\ref{eq:jb:BlochEq}) on $SO(3)$ and to \cite{LeonardKrishna} and  \cite{brockett2}
 for information on the general application of area rules in two-input control systems.  As detailed in \cite{JrShinLi}, systems of the form (\ref{eq:jb:BlochEq}) can be used to guide the design of rf pulses for quantum control experiments.  
While (\ref{eq:jb:BlochEq}) does not have the same natural bilinear input-output structure as (\ref{eq:wong:BI}) the approximate relationship between the two system responses to closed curve inputs augurs well for potential applications of the computability results to be established in the remainder of the paper.
\end{example}

With such broader applications of B-H systems in mind, we confine our attention in the remainder of the paper to studying the response of (\ref{eq:wong:BI}) to those closed input curves $(u(t),v(t))$ that give rise to closed curves $(x(t),y(t))$.  Recall that $\cU=\{ u_1,\ldots, u_m \}$ and $\cV=\{ v_1,\ldots, v_n \},$
represent the sets of the control actions.  We assume the controls are elements in $\mathcal{L}$, a closed subspace of $L^2[0,1]$ consisting of functions that can be represented by the following type
of Fourier series with square summable coefficients: 
\begin{equation}
\begin{array}{lll}
u_i(t)= \sqrt{2} \sum_{k=1}^\infty \left[a_{i,2k-1}  \sin(2\pi kt) +  a_{i,2k}  \cos(2\pi kt) \right], \\
v_j(t)= \sqrt{2} \sum_{k=1}^\infty \left[ -b_{j,2k-1}  \cos(2\pi kt) + b_{j,2k}  \sin(2\pi kt) \right].
\end{array}
\label{eq:Fourier}
\end{equation}
This restriction is not essential to the investigation reported in this paper, but it
allows connection to the results presented in \cite{WB} and \cite{CDC09}.
Note that $\mathcal{L}$ contains continuous periodic functions with zero mean and period one.
From basic orthogonality properties of the sine and cosine functions, one can show:
\begin{lemma}
\label{lemma:1}
If the controls $u_i$ and $v_j$ are used, then at time $t=1$,
\begin{equation}
\begin{array}{lll}
z(1)&=&  \sum_{k=1}^\infty \dfrac{a_{i,2k-1}b_{j,2k-1}}{\pi k} + \sum_{k=1}^\infty \dfrac{a_{i,2k}b_{j,2k}}{\pi k} .
\end{array}
\end{equation}
Moreover,
\begin{equation}
\int_0^1 u_i^2(t)dt =  \sum_{k=1}^\infty (a_{i,2k-1}^2+a_{i,2k}^2),
\end{equation}
\begin{equation}
\int_0^1 v_j^2(t)dt =  \sum_{k=1}^\infty (b_{j,2k-1}^2+b_{j,2k}^2).
\end{equation}
\end{lemma}

The proof is a straightforward calculation and is omitted.
Define matrices $\mathbf{U_B}$, $\mathbf{V_B}$, and $\mathbf{F_B}$ as follows:
\begin{equation}
\mathbf{U_B}=
\left[
\begin{array}{llll}
a_{1,1} & a_{1,2} & a_{1,3} & \ldots \vspace{-0.2cm}\\
a_{2,1} & a_{2,2} & a_{2,3} & \ldots \vspace{-0.2cm}\\
\vdots & \vdots & \vdots & \ddots \vspace{-0.2cm}\\
a_{m,1} & a_{m,2} & a_{m,3} & \ldots
\end{array}
\right],
\end{equation}
\begin{equation}
\mathbf{V_B}=
\left[
\begin{array}{llll}
b_{1,1} & b_{1,2} & b_{1,3} & \ldots \vspace{-0.2cm}\\
b_{2,1} & b_{2,2} & b_{2,3} & \ldots \vspace{-0.2cm}\\
\vdots & \vdots & \vdots & \ddots \vspace{-0.2cm}\\
b_{n,1} & b_{n,2} & b_{n,3} & \ldots 
\end{array}
\right],
\end{equation}
\begin{equation}
\mathbf{F_B} \equiv
\frac{1}{\pi}\left[
\begin{array}{lllllll}
1 & 0 & 0 & 0 & 0 & 0 & \ldots \vspace{-0.2cm}\\
0 & 1 & 0 & 0 & 0 & 0 & \ldots \vspace{-0.2cm}\\
0 & 0 & 2^{-1} & 0 & 0 & 0 &  \ldots \vspace{-0.2cm}\\
0 & 0 & 0 & 2^{-1} & 0 & 0 &  \ldots \vspace{-0.2cm}\\
0 & 0 & 0 & 0 & 3^{-1} & 0 & \ldots \vspace{-0.2cm}\\
0 & 0 & 0 & 0 & 0 & 3^{-1} & \ldots \vspace{-0.2cm}\\
\vdots & \vdots & \vdots & \vdots & \vdots & \vdots & \ddots \vspace{-0.2cm}\\
\end{array}
\right].
\label{eq:FB}
\end{equation}

\begin{lemma}
The output of the B-H system is given by 
$\mathbf{U_B} \mathbf{F_B} \mathbf{V_B}^T.$
Moreover,
\begin{equation}
\sum_{i=1}^m \int_0^1 u_i^2(t)dt =  tr \mathbf{U_B}  \mathbf{U_B}^T, \;\;
\sum_{i=1}^n \int_0^1 v_i^2(t)dt =  tr \mathbf{V_B}  \mathbf{V_B}^T.
\end{equation}
\end{lemma}

The proof is straightforward and is omitted.  It should be noted that the weight assignment
and the indexing in (\ref{eq:Fourier}) are chosen to allow 
the matrices $\mathbf{U_B}, \mathbf{V_B}$, and $\mathbf{F_B}$ to assume these simple representations.
The control cost defined in (\ref{eq:control cost}) can be rewritten as
\begin{equation}
I(\mathcal{U},\mathcal{V}) = \left(\frac{1}{m} tr \mathbf{U_B} \mathbf{U_B}^T +
\frac{1}{n}tr \mathbf{V_B} \mathbf{V_B}^T 
\right).
\label{eq:P2}
\end{equation}

These results can be generalized to any system
where the input-output mapping, $F$, 
is a bounded functional on $\mathcal{L} \otimes \mathcal{L}$.
Let $\{e_1, e_2, \ldots \}$ and $\{f_1, f_2, \ldots \}$ be orthonormal bases (possibly the same) for $\mathcal{L}$ 
with respect to the standard inner product on $L^2[0,1]$.
Let $L$ be the order of $\mathcal{L}$, which could be infinity.
If $u=\sum_{i=1}^L r_i e_i$ and $v=\sum_{j=1}^L s_j f_j$,
by the bounded bilinear 

\begin{equation}
F(u, v) = F(\sum_{i=1}^L r_i e_i, \sum_{j=1}^L s_j f_j)=\sum_{i,j=1}^L r_is_j F(e_i,f_j).
\end{equation}
We can represent $u_i \in \mathcal{U}$ and $v_j \in \mathcal{V}$ as $L$ dimensional row vectors by means of their coefficients with respect
to these orthonormal bases.  Let
$\mathbf{U}$ be the $m$ by $L$ matrix whose $i$-th row is the vector representation of $u_i$ and
$\mathbf{V}$ be the $n$ by $L$ matrix whose $j$-th row is the vector representation of $v_j$.
Under these bases, $F$ has an $L$ by $L$ dimensional matrix representation:
$\mathbf{F}\equiv
\left[
F(e_i,f_j)
\right];$ that is,\ $F(e_i,f_j)=F_{ij}$.

\begin{proposition}
The condition that the targets are achievable can be expressed by the equation:
\begin{equation}
\mathbf{H}=\mathbf{U}\mathbf{F}\mathbf{V}^T.
\label{eq:P3}
\end{equation}
\end{proposition}

The proof is straightforward and is omitted.

As usual, define the rank of $\mathbf{F}$ to be the number of its independent columns (or rows).
The rank of ${\bf F}$ is said to be {\em infinite} if there is no finite subset of columns (rows) in terms of which all columns (rows) can be expressed as linear combinations.  One can show via arguments presented in
Appendix A that the rank is independent of the bases chosen.
Hence, we can speak of the rank of a bilinear map without ambiguity.

\begin{proposition}
There exists a single round protocol that realizes an $m$-by-$n$ target matrix, $\mathbf{H}$, of rank $k$, 
if and only if the rank of the bilinear input-output mapping $F$ is at least $k$.
\end{proposition}

\noindent
\textit{Proof}: Let $\mathbf{H}_k$ be a full rank $k$-by-$k$ sub-matrix of $\mathbf{H}$. 
If there is a single round protocol to realize $\mathbf{H}$, we can select controls from the solution protocol
to obtain a realization for $\mathbf{H}_k$.  Conversely, if a single round protocol exists for $\mathbf{H}_k$,
we can extend it to obtain a protocol for realizing $\mathbf{H}$ by adding control functions to the
control sets according to the additional choices allowed; these added controls can be constructed as
linear combinations of the original controls used in realizing $\mathbf{H}_k$.  
Hence, we can assume without loss
of generality that $m=n=k$ and $\mathbf{H}$ is a $k$-by-$k$ full rank matrix.

Suppose a single round protocol exists for such a target matrix.  The solutions consist of
$k$ controls for Alice and $k$ for Bob.  We can construct an orthonormal basis for
$\mathcal{L}$ so that the controls used by Alice are spanned
by the first $k$ basis elements.   Construct similarly a basis for Bob.
Represent $F$ in terms of these bases and let $\mathbf{\hat{F}}$
denote the restriction of $F$ to the first $k$ basis elements.
Then, $\mathbf{\hat{F}}$ is a $k$-by-$k$ matrix and the equation
\begin{equation}
\mathbf{H}=\mathbf{U}\mathbf{\hat{F}}\mathbf{V}^T
\label{eq:P4}
\end{equation}
has a solution.  It follows that the rank of $\mathbf{\hat{F}}$, and hence, $F$ is at least $k$.

Conversely, if there exists a $k$-by-$k$ matrix $\mathbf{\hat{F}}$ of full rank representing the
restriction of $F$ to some control subspaces, equation (\ref{eq:P4}) always has a solution for
any $k$-by-$k$ matrix, $\mathbf{H}$.			\hfill $\blacksquare$

\vspace{.5cm}

The previous proposition implies that the existence of a target realizing single round protocol
depends on the rank of the bilinear input-output mapping and the rank of the target matrix.  
For a bilinear input-output mapping of finite rank,
there exists target matrices that cannot be realized without communication between the agents.
For a bilinear input-output mapping of infinite order, any finite dimensional target matrix can be
realized by a single round protocol.   

Given a matrix representation, $\mathbf{F}$, for $j=1,2, \ldots$,
define its $j$-{\em dimensional leading principal minors} by:
\begin{equation}
\mathbf{F}_l=\left[
\small
\begin{array}{llll}
F_{1,1} &  \ldots & F_{1,j} \vspace{-0.2cm}\\
\vdots & \ddots & \vdots \vspace{-0.2cm}\\
F_{l,j} & \ldots & F_{j,j}
\end{array}
\right].
\label{eq:JB:Leading}
\end{equation}

\begin{definition}
A matrix representation of a bilinear map is {\it regular} if 
for all positive integers $j$ less than or equal to its rank, its $j$-dimensional leading principal minors are nonsingular.
\end{definition}

\begin{definition}
A matrix representation of a bilinear map is {\it strongly regular} if it is finite
dimensional and regular or if it is in
diagonal form such that for all positive integers $j$
the diagonal elements satisfy the ordering
\begin{equation}
F_{j,j} \geq F_{j+1,j+1} > 0.
\label{eq:strongreg}
\end{equation}
\end{definition}
It is easy to find examples of finite dimensional regular or strongly regular
matrix representations.  For the B-H system, the input-output mapping that takes $(u,v)$ to $z(1)$ by means of
(\ref{eq:wong:BI}) has a matrix representation $\mathbf{F}_\bB$ described by (\ref{eq:FB})
with respect to the orthonormal basis,
\begin{align}
\mathcal{B}_0 = \{\sqrt{2} \sin(2\pi t), \sqrt{2}\cos(2\pi t),\sqrt{2}\sin(4\pi t), \nonumber\\
\sqrt{2}\cos(4\pi t), \ldots \}.
\label{eq:JB:OurBasis}
\end{align}
The matrix is diagonal of infinite rank and provides an infinite dimensional strongly regular representation of $F$.  

To prepare for subsequent discussions
we recall some basic ideas on singular values.   A $p$-by-$q$ matrix, 
$\mathbf{M}$, has $p$ singular values while $\mathbf{M}^T$ has $q$ singular values.  It is well-known that the nonzero singular
values of the two matrices are identical (see \cite{mo} for details).
We denote by $\sigma_i(\mathbf{M})$ the $i$-th singular value of $\mathbf{M}$
under the ordering
\begin{equation}
\sigma_1(\mathbf{M}) \geq \sigma_2(\mathbf{M}) \ldots \geq \sigma_p(\mathbf{M}).
\end{equation}
If $\mathbf{M}$ is a $p$-by-$p$ symmetric matrix, represent by $\lambda_i(\mathbf{M})$ the $i$-th eigenvalue under the ordering
\begin{equation}
\lambda_1(\mathbf{M}) \geq \lambda_2(\mathbf{M}) \ldots \geq \lambda_p(\mathbf{M}).
\end{equation}

\begin{lemma}
Consider a $k$-by-$k$ matrix, $\mathbf{M}$, with a matrix decomposition so that
\begin{equation}
\mathbf{M}=\left[
\begin{array}{ll}
\mathbf{M}_{1,1} & \mathbf{M}_{1,2}\\
\mathbf{M}_{2,1} & \mathbf{M}_{2,2}
\end{array}
\right],
\end{equation}
where $\mathbf{M}_{1,1}$ is $l$-by-$l$,
$\mathbf{M}_{1,2}$ is $l$-by-$(k-l)$, $\mathbf{M}_{2,1}$ is $(k-l)$-by-$l$,
and $\mathbf{M}_{2,2}$ is $(k-l)$-by-$(k-l)$.  Then, for $i=1, \ldots, l$
\begin{equation}
\sigma_i(\mathbf{M}) \geq \sigma_i(\mathbf{M}_{1,1}).
\end{equation}
\end{lemma}

\noindent
\textit{Proof}:  Consider the following positive semi-definite matrix:
\begin{align}
& \mathbf{MM}^T \nonumber \\
&=\left[
\begin{array}{ll}
\mathbf{M}_{1,1} \mathbf{M}_{1,1}^T + \mathbf{M}_{1,2}\mathbf{M}_{1,2}^T
& \mathbf{M}_{1,1}\mathbf{M}_{2,1}^T + \mathbf{M}_{1,2}\mathbf{M}_{2,2}^T\\
\mathbf{M}_{2,1}\mathbf{M}_{1,1}^T + \mathbf{M}_{2,2}\mathbf{M}_{1,2}^T
& \mathbf{M}_{2,1}\mathbf{M}_{2,1}^T + \mathbf{M}_{2,2}\mathbf{M}_{2,2}^T
\end{array}
\right].
\end{align}
It follows from Fischer's Minimax Theorem (page 510, A.1.c \cite{mo}) that for
$i=1, \ldots, l$
\begin{equation}
\lambda_i(\mathbf{MM}^T) \geq \lambda_i(\mathbf{M}_{1,1}\mathbf{M}_{1,1}^T + \mathbf{M}_{1,2}\mathbf{M}_{1,2}^T).
\end{equation}
By means of a result of Loewner (page 510, A.1.b \cite{mo}):
\begin{equation}
\lambda_i(\mathbf{M}_{1,1}\mathbf{M}_{1,1}^T + \mathbf{M}_{1,2}\mathbf{M}_{1,2}^T)
\geq \lambda_i(\mathbf{M}_{1,1}\mathbf{M}_{1,1}^T ).
\end{equation}
The result then follows by combining these two inequalities.   \hfill $\blacksquare$

We conclude this section by noting that if a matrix representation, $\bF$, is
finite dimensional, then its singular values are well-defined and we use
$\sigma_i(\mathbf{F})$ to represent the $i$-th largest singular value.
One can extend this concept to the infinite dimensional case in the following way.
For any matrix $\bF$ and corresponding sequence of principal minors $(\bF_1, \bF_2,\dots )$ (\ref{eq:JB:Leading}), it follows from the preceding discussion that 
\begin{equation}
\sigma_i(\mathbf{F}_l) \leq \sigma_i(\mathbf{F}_{l+1}).
\end{equation}
On the other hand, it is shown in Appendix B that
$\sigma_i(\mathbf{F}_l) \leq \|F\|$
for all $i$.  Hence the limit
$\lim_{l \rightarrow \infty} \sigma_i(\mathbf{F}_l)$
exists and is finite.  Denote this limit by $\sigma_i(\mathbf{F})$.
If $\bF$  is a strongly regular infinite dimensional representation then clearly
for all integers $i$ and $l$, $
\sigma_i(\mathbf{F}) = \sigma_i(\mathbf{F}_{l}) .
$

\section{MATRIX REPRESENTATION OF THE OPTIMIZATION PROBLEM}\setcounter{equation}{0}

The control cost of $\mathcal{U}$ and $\mathcal{V}$ defined in (\ref{eq:control cost})
can be rewritten in matrix form as
\begin{equation}
\frac{1}{m} tr \mathbf{U} \mathbf{U}^T +
\frac{1}{n}tr \mathbf{V} \mathbf{V}^T. 
\end{equation}

Our goal of finding the optimal control to realize a given target matrix can now
be represented as a matrix optimization problem.   One is also interested in
approximate solutions restricted to finite dimensional control subspaces.
In particular, for any positive integer $l$, satisfying $l \geq \max(m,n)$,
consider the following optimization problem involving the
$l$-by-$l$ leading principal minor of  $\mathbf{F}$, $\mathbf{F}_l$:

\noindent \textbf{Optimization Problem} $(\mathbf{H}, \mathbf{F}_l)$:
Let $\mathbf{U}$ and $\mathbf{V}$ be $m$-by-$l$ and $n$-by-$l$ matrices respectively.
The optimization problem is defined by:
\begin{equation}
\min_{\mathbf{U},\mathbf{V}}  \left( \frac{1}{m} tr \mathbf{U} \mathbf{U}^T +
\frac{1}{n}tr \mathbf{V} \mathbf{V}^T \right)
\end{equation}
subject to the constraint:
\begin{equation}
\mathbf{H}=\mathbf{U}\mathbf{F}_l\mathbf{V}^T.
\end{equation}

While it is not clear whether optimal sequences of controls for the above problem always exist, the infimum cost, denoted by $\hat{C}_F(\mathbf{H})$, is always well-defined if the target matrix is realizable.

One can formulate a slightly more general version of this optimization problem by allowing the
weights in the cost function to be arbitrary positive integers:

\noindent \textbf{Generalized Optimization Problem} $(\mathbf{H}, \mathbf{F}_l; p, q)$:
Let $\mathbf{H}$ be an $m$-by-$n$ target matrix,
$\mathbf{F}_l$ be an $l$-by-$l$ leading principal minor of  $\mathbf{F}$, with  $l \geq max(m,n)$,
$p$ and $q$ be arbitrary positive integers.  The optimization problem is defined by:
\begin{equation}
\min_{\mathbf{U},\mathbf{V}} \left( \frac{1}{p} tr \mathbf{U} \mathbf{U}^T +
\frac{1}{q}tr \mathbf{V} \mathbf{V}^T \right)
\end{equation}
subject to the condition
\begin{equation}
\mathbf{H}=\mathbf{U}\mathbf{F}_l\mathbf{V}^T.
\label{eq:P5}
\end{equation}

There is an important connection between these two classes of problems.
Given an $m$-by-$n$ target matrix, $\mathbf{H}$, one can obtain another target matrix by
appending rows of zeros and columns of zeros to obtain an $l$-by-$l$ matrix, with $l \geq n, l \geq m$, so that
\begin{equation}
\mathbf{\tilde{H}}=\left[
\begin{array}{ll}
\mathbf{H} & \mathbf{0}_{m,l-n}\\
\mathbf{0}_{l-m,n} & \mathbf{0}_{l-m,l-n}
\end{array}
\right],
\end{equation}
where $\mathbf{0}_{i,j}$ is an $i$-by-$j$ matrix with all zeros.

Now consider the Generalized Optimization Problem $(\mathbf{\tilde{H}, F}_l; m, n)$.
Any optimal solution to the Generalized Optimization Problem must satisfy the property:
\begin{equation}
u_i=0, \; v_j=0
\end{equation}
for $i>m$ and $j>n$.  Otherwise, a lower cost solution can be obtained by substituting with these
zero controls.   From this, one can conclude that the solution is also optimal for the lower
dimensional Optimization Problem, $(\mathbf{H,F}_l)$.
Conversely, an optimal solution to the latter problem can be extended to an optimal solution to
the Generalized Optimization Problem $(\mathbf{\tilde{H},F}_l;m,n)$.
Hence, by using the Generalized Optimization Problem formulation, we can assume without loss
of generality that the target matrices are square matrices with the same dimensions as $\mathbf{F}_l$.

\section{SINGLE ROUND PROTOCOLS: MINIMUM ENERGY CONTROL}\setcounter{equation}{0}

One of the key results in this paper is summarized by the following theorem.

\begin{theorem}
\label{thm:1}
Consider a bounded bilinear input-output mapping, $F$, with a regular matrix representation $\mathbf{F}$ with rank $s$.
Let $\mathbf{H}$ be an $m$-by-$n$ target matrix, such that
$s \geq max(m,n)$. 
The infimum control cost of any single round protocol that realizes $\mathbf{H}$ is given by:
\begin{equation}
\hat{C}_F(\mathbf{H}) =
\frac{2}{\sqrt{mn}}
\sum_{k=1}^{\min(m,n)} \sigma_{k}(\mathbf{H})/\sigma_{k}(\mathbf{F}).
\label{eqn:thm1}
\end{equation}
If the matrix representation is strongly regular, there exists a single round protocol that achieves this infimum control cost.
\end{theorem}

Before proving Theorem \ref{thm:1}, we present a  corollary that specializes the result to the B-H system (\ref{eq:wong:BI}). 
First note that the diagonal entries of $\mathbf{F_B}$ in (\ref{eq:FB}) can be represented as $(\lceil k/2 \rceil \pi)^{-1}$, and this leads to the following result.

\begin{corollary}
\label{cor:1}
Consider the input-output system (\ref{eq:wong:BI}) and an $m$-by-$n$ target matrix $\mathbf{H}$.
For single round protocols that realize $\mathbf{H}$ the infimum control cost is given by
\begin{equation}
\hat{C}_{\bF_\bB}(\mathbf{H}) =\frac{2\pi}{\sqrt{mn}}
\sum_{k=1}^{\min(m,n)} \lceil k/2 \rceil  \sigma_{k}(\mathbf{H}),\label{eq:thm5.1}
\end{equation}
and this is achieved by Alice choosing $m$ controls and Bob choosing $n$ controls
from  the space spanned by the basis $\mathcal{B}_0$ defined in (\ref{eq:JB:OurBasis}).
\end{corollary}

To prove the main theorem, we first establish a proposition in which we prove the result for
a square target matrix $\mathbf{H}$ and for controls restricted to finite dimensional subspaces.

\begin{proposition}
\label{prop:1}
Consider an invertible matrix $\mathbf{F}_l$ and an $l$-by-$l$ target matrix $\mathbf{H}$ with rank $r$.
The minimum control cost for the Generalized Optimization Problem $(\mathbf{H}, \mathbf{F}_l;p,q)$
is achievable and is given by the formula,
\begin{equation}
\hat{C}_{\mathbf{F}_l}(\mathbf{H})=
\frac{2}{\sqrt{pq}}
\sum_{k=1}^{r} \sigma_{k}(\mathbf{H})/\sigma_{k}(\mathbf{F}_l).
\label{eq:prop1.1}
\end{equation}
\end{proposition}

\noindent
\textit{Proof}:
To prove the proposition, first of all we show that the right-hand-side of (\ref{eq:prop1.1}) can be achieved.
Let $\Pi$ be the orthogonal matrix that puts $\bF_l \bF_l^T$ into the following diagonal form:
\begin{equation}
\Pi^T \bF_l\bF_l^T  \Pi =
\left[
\begin{array}{ccc}
\sigma_1^2(\mathbf{F}_l) & \ldots   &  0 \\
\vdots & \ddots  &\vdots   \\
0  &  \ldots &   \sigma_l^2(\mathbf{F}_l)
\end{array}
\right].
\label{eq:prop1.2}
\end{equation}
Define $\mathbf{\tilde{U}}=\Pi^T \mathbf{U} \Pi$ and $\mathbf{\tilde{V}}=\Pi^T \mathbf{V} \Pi$.
Then $(\mathbf{U}, \mathbf{V})$ is a solution to
\begin{equation}
\mathbf{H}=\mathbf{U}\mathbf{F}_l\mathbf{V}^T
\label{eq:prop1.3}
\end{equation}
if and only if $(\mathbf{\tilde{U}},\mathbf{\tilde{V}})$
is a solution to the equation
\begin{equation}
\Pi^T \bH \Pi = \mathbf{\tilde{U}} \Pi^T \bF_l \Pi \mathbf{\tilde{V}}^T.
\end{equation}
Since the cost of $(\mathbf{U}, \mathbf{V})$ and $(\mathbf{\tilde{U}},\mathbf{\tilde{V}})$ are identical, we can assume
without loss of generality that $\bF_l \bF_l^T$ is in the diagonal form (\ref{eq:prop1.2}).
Let $\mathbf{\Theta}$ be the orthogonal matrix that diagonalizes $\mathbf{H}^T\mathbf{H}$ so that
\begin{equation}
\mathbf{\Theta} \mathbf{H}^T \mathbf{H} \mathbf{\Theta}^T =
\left[
\begin{array}{cccccc}
\sigma_1^2(\mathbf{H}) && \ldots   &  && 0  \vspace{-0.2cm}\\
&\ddots&&&& \vspace{-0.2cm}\\
\vdots&&\sigma_r^2(\mathbf{H})&&&\vdots \vspace{-0.2cm}\\
&&& 0&& \vspace{-0.2cm}\\
&&&& \ddots  &   \vspace{-0.2cm}\\
0 & &  \ldots && &  0
\end{array}
\right],
\label{eq:THETA}
\end{equation}
and let
\begin{align}
&\mathbf{R_{\delta}} =
\left( \frac{q}{p} \right)^{1/4}\nonumber\\
 & \cdot \left[
\begin{array}{cccccc}
\sqrt{\sigma_1(\mathbf{H})\sigma_1(\mathbf{F}_l)} && \ldots   &  && 0  \vspace{-0.2cm}\\
&\ddots&&&& \vspace{-0.2cm}\\
\vdots&&\sqrt{\sigma_r(\mathbf{H})\sigma_r(\mathbf{F}_l)}&&&\vdots \vspace{-0.2cm}\\
&&& \delta&& \vspace{-0.2cm}\\
&&&& \ddots  &    \vspace{-0.2cm}\\
0 & &  \ldots && &  \delta
\end{array}
\right]
\label{eq:R}
\end{align}
for a small $\delta>0$.  In equation (\ref{eq:THETA}) the lower right-hand-side zero-diagonal block is absent if $r=l$.
Similarly for equation (\ref{eq:R}), the lower right-hand-side $\delta$-diagonal block is absent if $r=l$.   Define
\begin{equation}
\mathbf{U}_\delta=\mathbf{H} \mathbf{\Theta} \mathbf{R_{\delta}}^{-1}, \;
\mathbf{V}_\delta^T=\mathbf{F}_l^{-1} \mathbf{R_{\delta}} \mathbf{\Theta}^T,
\end{equation}
we obtain a solution to (\ref{eq:prop1.3}).  
By direct computation, it follows that
\begin{align}
\frac{1}{p} tr \mathbf{U}_\delta \mathbf{U}_\delta^T &= \frac{1}{p}  tr \mathbf{H}\mathbf{\Theta}
\mathbf{R}_\delta^{-2} \mathbf{\Theta}^T \mathbf{H}^T \\
&=  \frac{1}{p} tr \mathbf{R}_\delta^{-2} \mathbf{\Theta}^T \mathbf{H}^T \mathbf{H} \mathbf{\Theta}\\
&=  \frac{1}{\sqrt{pq}} \sum_{k=1}^{r} \sigma_{k}(\mathbf{H})/\sigma_{k}(\mathbf{F}_l).
\label{eq:UU}
\end{align}
Since entries of $\mathbf{U}_\delta$ are affine functions of $1/\delta$, equation (\ref{eq:UU})
implies $\mathbf{U}_\delta$
is independent of $\delta$.   We express this by writing $\mathbf{U}_\delta\equiv\mathbf{U}_0$.  It is clear that
$\lim_{\delta \rightarrow 0} \mathbf{V}_\delta$
exists.  Denote it by $\mathbf{V}_0$.
Then,
\begin{align}
\frac{1}{q} tr \mathbf{V}_\delta \mathbf{V}_\delta^T & =  \frac{1}{q} tr \mathbf{\Theta} \mathbf{R}_\delta
(\mathbf{F}_l\mathbf{F}_l^T)^{-1} \mathbf{R}_\delta \mathbf{\Theta}^T\\
&= \frac{1}{q} tr \mathbf{R}_\delta(\mathbf{F}_l\mathbf{F}_l^T)^{-1} \mathbf{R}_\delta \\
&= \frac{1}{q} tr \mathbf{R}_\delta^2 (\mathbf{F}_l\mathbf{F}_l^T)^{-1}\\
& = \frac{1}{\sqrt{pq}} \sum_{k=1}^{r} \frac{\sigma_{k}(\mathbf{H})}{\sigma_{k}(\mathbf{F}_l)} + \frac{\delta^2}{\sqrt{pq}} \sum_{k=r+1}^{l} \frac{1}{\sigma_{k}^2(\mathbf{F}_l)}.
\label{eq:VV}
\end{align}
Taking the limit as $\delta\rightarrow 0$,
\begin{equation}
\frac{1}{p}tr\mathbf{U}_0\mathbf{U}_0^T + \frac{1}{q}tr \mathbf{V}_0\mathbf{V}_0^T
= \frac{2}{\sqrt{pq}} \sum_{k=1}^{r} \sigma_{k}(\mathbf{H})/\sigma_{k}(\mathbf{F}_l),
\label{eq:5.12}
\end{equation}
proving that the right-hand-side of (\ref{eq:prop1.1}) can be achieved.

To complete the proof, we want to show  the right-hand-side of (\ref{eq:prop1.1}) is a lower bound for all
single round protocols realizing $\mathbf{H}$.   We show this first for invertible ${\mathbf H}$.  In this case, all solutions
$(\mathbf{U},\mathbf{V})$ to the equation (\ref{eq:prop1.3})
consist of invertible matrices and for every invertible $\mathbf{V}$ there is a unique matrix ${\mathbf U}$ that solves (\ref{eq:prop1.3}).
By applying polar decomposition to $\mathbf{F}_l\mathbf{V}^T$ we obtain
\begin{equation}
\mathbf{V}^T =\mathbf{F}_l^{-1}\mathbf{R}\mathbf{\Theta}^T,
\end{equation} 
for an orthogonal matrix, $\mathbf{\Theta}$, and a non-singular symmetric matrix, $\mathbf{R}$.
These two matrices can be regarded as free variables and we can parametrize the solution space
to (\ref{eq:prop1.1}) by $\mathbf{R}$ and $\mathbf{\Theta}$.
Specifically, if we let $\mathbf{U}=\mathbf{H}\mathbf{\Theta}\mathbf{R}^{-1}$, then the control cost is
\begin{align}
&\quad \frac{1}{p}tr \mathbf{U} \mathbf{U}^T +
\frac{1}{q} tr \mathbf{V} \mathbf{V}^T \nonumber \\
&=
\frac{1}{p} tr \mathbf{R}^{-2} \mathbf{\Theta}^T \mathbf{H}^T \mathbf{H} \mathbf{\Theta}
+ \frac{1}{q} tr \mathbf{R}^2 (\mathbf{F}_l\mathbf{F}_l^T)^{-1}.
\end{align}

For any two $l$-by-$l$ positive semidefinite symmetric matrices, 
$\mathbf{P}$ and $\mathbf{Q}$
\begin{equation}
tr \mathbf{PQ} = \sum_{k=1}^l \lambda_k(\mathbf{PQ}) \geq \sum_{k=1}^l \lambda_k(\mathbf{P}) \lambda_{l-k+1}(\mathbf{Q}).
\end{equation}
(See, for instance, p.249 of \cite{mo}.)  Therefore,
\begin{align}
tr \mathbf{R}^2 (\mathbf{F}_l\mathbf{F}_l^T)^{-1} 
&\geq \sum_{k=1}^l \lambda_k^2(\mathbf{R}) \sigma_{l-k+1}^2(\mathbf{F}_l^{-1})\\
&= \sum_{k=1}^l \lambda_k^2(\mathbf{R})/\sigma_{k}^2(\mathbf{F}_l).
\end{align}
Similarly,
\begin{equation}
\begin{array}{lll}
tr \mathbf{R}^{-2} \mathbf{\Theta}^T \mathbf{H}^T \mathbf{H} \mathbf{\Theta}
\geq  \sum_{k=1}^l \sigma_k^2(\mathbf{H}) / \lambda_k^{2}(\mathbf{R}).
\end{array}
\end{equation}
Hence,
\begin{equation}
I(\mathcal{U},\mathcal{V})  \geq \dfrac{1}{p} \sum_{k=1}^l \sigma_k^2(\mathbf{H}) / \lambda_k^{2}(\mathbf{R}) +
\dfrac{1}{q}\sum_{k=1}^l \lambda_k^2(\mathbf{R})/\sigma_{k}^2(\mathbf{F}_l) .
\label{eq:lowerbound}
\end{equation}
One can minimize the right-hand-side of (\ref{eq:lowerbound}) by considering
\begin{equation}
\min \sum_{k=1}^l \left[ \dfrac{1}{p}\dfrac{\sigma_k^2(\mathbf{H})}{t_k} 
+ \dfrac{1}{q}\dfrac{t_k}{\sigma_{k}^2(\mathbf{F}_l) } \right],
\end{equation}
subject to the constraint
\begin{equation}
t_1 \geq t_2 \ldots \geq t_l > 0.
\label{eq:t1}
\end{equation}
Via calculus, the unconstrained optimal solution to (\ref{eq:lowerbound}) is shown to be
\begin{equation}
t_k=\sigma_k(\mathbf{H})\sigma_k(\mathbf{F}_l)\sqrt{\frac{q}{p}},
\end{equation}
which also satisfies the condition in (\ref{eq:t1}).  It follows that
\begin{equation}
\hat{C}_{\mathbf{F}_l}(\mathbf{H}) \geq \frac{2}{\sqrt{pq}}
\sum_{k=1}^l \sigma_k(\mathbf{H})/\sigma_{k}(\mathbf{F}_l).
\label{ineq:prop1}
\end{equation}
By (\ref{eq:5.12}) the right-hand-side of (\ref{ineq:prop1}) can be achieved, hence the inequality is an equality and the proposition holds for invertible $\bf H$, that is $r=l$.

We want to show the inequality (\ref{ineq:prop1}) also holds when $\bf H$ is not full rank.
Suppose $\hat{\bf U}$ and $\hat{\bf V}$ give an optimal solution to $({\bf H},{\rm F}_l;p,q)$.
Define
\begin{equation}
\mathbf{U} (\epsilon)=\mathbf{\hat{U}}+\epsilon \mathbf{I}, \;
\mathbf{V} (\epsilon)=\mathbf{\hat{V}}+\epsilon \mathbf{I},
\end{equation}
where $\mathbf{I}$ is the $l$-by-$l$ identity matrix.  Other than at most a finite set of values, all
these matrices are invertible.    Without loss of generality, assume that for some $a>0$ and all $0<\epsilon \leq a$,
the matrices $\mathbf{U}(\epsilon)$ and $\mathbf{V}(\epsilon)$ are invertible
and restrict $\epsilon$ in subsequent discussion to such an interval.   Let
$\mathcal{U}(\epsilon)$ denote the set of controls corresponding to $\mathbf{U}(\epsilon)$
and similarly,  $\mathcal{V}(\epsilon)$ denote the sets of controls corresponding to $\mathbf{V}(\epsilon)$.
Define
\begin{equation}
\mathbf{G}(\epsilon)=\mathbf{U}(\epsilon) \mathbf{F}_l \mathbf{V}(\epsilon)^T=
\mathbf{H}+\epsilon(\mathbf{\hat{U}}\mathbf{F}_l+\mathbf{F}_l \mathbf{\hat{V}}^T)+\epsilon^2 \mathbf{F}_l.
\end{equation}
For $0<\epsilon \leq a$, the matrices $\mathbf{G} (\epsilon)$ are invertible.  Hence,

\begin{align}
I(\mathcal{U},\mathcal{V})  & = \frac{1}{p} tr \mathbf{\hat{U}} \mathbf{\hat{U}}^T +
\frac{1}{q}tr \mathbf{\hat{V}} \mathbf{\hat{V}}^T \\
& = \lim_{\epsilon \rightarrow 0} \left( \frac{1}{p} tr \mathbf{U}(\epsilon) \mathbf{U}(\epsilon)^T +
\frac{1}{q}tr \mathbf{V}(\epsilon) \mathbf{V}(\epsilon)^T \right) \\
& \geq  \lim_{\epsilon \rightarrow 0}\hat{C}_{\mathbf{F}_l}(\mathbf{G}(\epsilon))  \label{eq:john:generalH1}\\
& = \lim_{\epsilon \rightarrow 0} \frac{2}{\sqrt{pq}}
\sum_{k=1}^{l} \sigma_{k}(\mathbf{G}(\epsilon))
/\sigma_k(\mathbf{F}_l) \label{eq:generalH1}\\
&=\frac{2}{\sqrt{pq}}  \sum_{k=1}^{l} \sigma_{k}(\mathbf{H})/\sigma_k(\mathbf{F}_l) \label{eq:generalH2}\\
&=\frac{2}{\sqrt{pq}}  \sum_{k=1}^{r} \sigma_{k}(\mathbf{H})/\sigma_k(\mathbf{F}_l) = \hat{C}_{\mathbf{F}_l}(\mathbf{H}) .
\end{align}
Note that the equality (\ref{eq:generalH1}) follows from the fact that the proposition holds for invertible target matrices, while
the equality (\ref{eq:generalH2}) follows from the continuity of the singular values as a function of
the matrix coefficients.  This proves the proposition.
\hfill $\blacksquare$

\noindent
\textit{Proof of Theorem \ref{thm:1}}: 
\rm If $\bF$ is finite dimensional, the theorem follows from Proposition
\ref{prop:1} if we take $\bF_l$ to be $\bF_s$.  Hence, assume that $\bF$ is infinite dimensional and 
let $\{e_1,e_2,\ldots \}$ and $\{f_1,f_2,\ldots \}$ be the bases corresponding to the representation of $\mathbf{F}$.  Let
$\mathcal{\hat{U}}=\{\hat{u}_1, \ldots, \hat{u}_m\}$ and
$\mathcal{\hat{V}}=\{\hat{v}_1, \ldots, \hat{v}_n\}$
be the controls in an optimal solution for the Optimization Problem ($\mathbf{H,F}$), with corresponding
matrices $\mathbf{\hat{U}}$ and $\mathbf{\hat{V}}$ respectively.   Let
\begin{equation}
\hat{u}_i=\sum_{j=1}^\infty a_{ij}e_j, \; \hat{v}_i=\sum_{j=1}^\infty b_{ij}f_j.
\end{equation}
Then
\begin{equation}
I(\mathcal{\hat{U}},\mathcal{\hat{V}})=
\dfrac{1}{m}\sum_{i=1}^m \sum_{j=1}^\infty a_{ij}^2 + \dfrac{1}{n}\sum_{i=1}^n \sum_{j=1}^\infty b_{ij}^2,
\label{eq:I}
\end{equation}
and
\begin{equation}
\mathbf{H}= \mathbf{\hat{U}} \mathbf{F} \mathbf{\hat{V}}^T.
\end{equation}

Define approximating control functions
\begin{equation}
\hat{u}_i^{(l)}=\sum_{j=1}^l a_{ij}e_j, \; \hat{v}_i^{(l)}=\sum_{j=1}^l b_{ij}f_j
\end{equation}
with corresponding $m$-by-$l$ and $l$-by-$n$ matrices
\begin{equation}
\mathbf{\hat{U}}_l=
\left[ 
\begin{array}{lll}
a_{1,1} & \ldots & a_{1,l} \\
\vdots & \ddots & \vdots \\
a_{m,1} & \ldots  & a_{m,l}
\end{array}
\right], \;
\mathbf{\hat{V}}_l^T=
\left[
\begin{array}{lll}
b_{1,1} & \ldots & b_{1,n} \\
\vdots & \ddots & \vdots \\
b_{l,1} & \ldots  & b_{l,n}
\end{array}
\right].
\end{equation}
Then, equation (\ref{eq:I}) can be rewritten as
\begin{align}
I(\mathcal{\hat{U}},\mathcal{\hat{V}})&= \frac{1}{m} tr \mathbf{\hat{U}} \mathbf{\hat{U}}^T +
\frac{1}{n}tr \mathbf{\hat{V}} \mathbf{\hat{V}}^T\\
&=
\lim_{l\rightarrow \infty} \left(
\frac{1}{m} tr \mathbf{\hat{U}}_l \mathbf{\hat{U}}_l^T +
\frac{1}{n} tr \mathbf{\hat{V}}_l \mathbf{\hat{V}}_l^T \right).
\end{align}
Define the $m$-by-$n$ approximate target matrix by
\begin{equation}
\mathbf{H}_l=\left( H_{ij}^{(l)} \right) = \mathbf{\hat{U}}_l \mathbf{F}_l \mathbf{\hat{V}}_l^T,
\end{equation}
where $\mathbf{F}_l$ is the $l$-th principal minor of $\mathbf{F}$.  Since $F$ is bounded, 
\begin{equation}
H_{ij}=F( \hat{u}_i, \hat{v}_j)=\lim_{l \rightarrow \infty} F(\hat{u}_i^{(l)},\hat{v}_j^{(l)})
=\lim_{l \rightarrow \infty}  H_{ij}^{(l)}.
\end{equation}
In matrix form:
\begin{equation}
\lim_{l \rightarrow \infty}  \mathbf{\hat{U}}_l \mathbf{F}_l \mathbf{\hat{V}}_l^T
=\lim_{l \rightarrow \infty} \mathbf{H}_l
=\mathbf{H}.
\label{eq:Hlimit}
\end{equation}
Since the singular value function is continuous on the space of $m$-by-$n$ matrices, for any integer $k, 0 \leq k \leq m$,
we have
\begin{equation}
\lim_{l \rightarrow \infty} \sigma_k(\mathbf{H}_l)= \sigma_k(\mathbf{H}).
\label{eq:singular}
\end{equation}

For any integer $l \geqslant \max(m,n)$, define the $l$-by-$l$ augmented matrix $\mathbf{\tilde{H}}_l$ by
\begin{equation}
\mathbf{\tilde{H}}_l= \left[
\begin{array}{ll}
\mathbf{H}_l & \mathbf{0}_{m,l-n} \\
\mathbf{0}_{l-m,n} & \mathbf{0}_{l-m,l-n}
\end{array} \right].
\end{equation}
Here, $\mathbf{0}_{ij}$ represents an $i$-by-$j$ zero matrix, (empty when one of the dimensions is zero.)
As argued before, the minimum control costs for the Optimization Problem
$(\mathbf{H}_l, \mathbf{F}_l)$ and the Generalized Optimization Problem
$(\mathbf{\tilde{H}}_l, \mathbf{F}_l,m,n)$ are identical.  By Proposition \ref{prop:1},

\begin{align}
\frac{1}{m} tr \mathbf{\hat{U}}_l \mathbf{\hat{U}}_l^T +
\frac{1}{n}tr \mathbf{\hat{V}}_l \mathbf{\hat{V}}_l^T 
& \geq  \hat{C}_{\mathbf{F}_l}(\mathbf{\tilde{H}}_l)\\
& = \dfrac{2}{\sqrt{mn}}
\sum_{k=1}^{l} \sigma_{k}(\mathbf{\tilde{H}}_l)/\sigma_{k}(\mathbf{F}_l).
\label{eq:thm}
\end{align}
Moreover, the last expression in (\ref{eq:thm}) can be realized by some control functions. 

The rank of $\mathbf{\tilde{H}}_l$ is equal to the rank of
$\mathbf{H}_l$, which is at most $\min(m,n)$, moreover, the first $\min(m,n)$
singular values of the two matrices are identical.   It follows that
for $k >\min(m,n)$
\begin{equation}
\sigma_k(\mathbf{\tilde{H}}_l)=0,
\end{equation}
and the last expression in (\ref{eq:thm}) is equal to
\begin{equation}
\frac{2}{\sqrt{mn}} \sum_{k=1}^{\min(m,n)} \sigma_{k}(\mathbf{H}_l)/\sigma_{k}(\mathbf{F}_l)
\end{equation}
and can be realized.
The first part of the theorem then follows from:
\begin{align}
\frac{1}{m} tr \mathbf{\hat{U}} \mathbf{\hat{U}}^T +
\frac{1}{n}tr \mathbf{\hat{V}} \mathbf{\hat{V}}^T
& \geq
\lim_{l\rightarrow \infty}  \dfrac{2}{\sqrt{mn}}
\sum_{k=1}^{\min(m,n)}
\sigma_{k}(\mathbf{H}_l)/\sigma_{k}(\mathbf{F}_l) \nonumber\\
& =  \dfrac{2}{\sqrt{mn}} \sum_{k=1}^{\min(m,n)} \sigma_{k}(\mathbf{H})/\sigma_{k}(\mathbf{F}). \label{eq:generalH4}
\end{align}

Since $\bF$ is diagonal, 
$\sigma_k(\mathbf{F}_{l})=\sigma_k(\mathbf{F})$
for all integers $k$ and $l$,
the solution to the Optimization Problem $(\mathbf{H}_l, \mathbf{F}_{l})$ achieves the control cost given by the lower bound in (\ref{eq:generalH4}).  This completes the proof of Theorem 5.1. 
\hfill $\blacksquare$

We conclude the discussion in this section by pointing out that there is an apparent non-symmetry in the solution constructed in
the proof of Proposition \ref{prop:1}.   Even if $\mathbf{H}$ and $\mathbf{F}$ are symmetric, the optimal solution may not be symmetric
in the sense that $\mathbf{U}$ and $\mathbf{V}$ may not be identical.  This would appear less surprising if one considers
the example where $\mathbf{H}$ is
\begin{equation}
\left[
\begin{array}{cc}
0  &  1  \\
1 &   0
\end{array}
\right],
\end{equation}
and $\mathbf{F}$ is the 2-by-2 identity matrix.  Having a symmetric solution to the optimization problem
would imply $\mathbf{H}$ is non-negative semidefinite which is obviously not the case.  

\section{COST OF DISTRIBUTED ACTION}\setcounter{equation}{0}

Results in the preceding section provide an explicit formula for computing the control cost in the absence of
any communication between the agents.  However, if information can be shared between the agents, controls with lower average cost of distributed action can be designed.   For the B-H system, if both agents have information on the other agent's choice, they can use the following control functions to realize the target, $H_{ij}$:
\begin{eqnarray}
u_{ij}(t) &=& sgn(H_{ij}) \sqrt{2\pi H_{ij}} \cos(2\pi t),\\
v_{ij}(t) &=& \sqrt{2\pi H_{ij}} \sin(2\pi t).
\end{eqnarray}
The control energy of such a protocol is 
$2\pi \vert H_{ij} \vert$.
Using the isoperimetric inequality \cite{Os}, one can show as in \cite{WB} that this is the minimum
control cost to realize the single target, $H_{ij}$.

Hence, if both agents have complete information of the other agent's choice,
the averaged control cost over all possible choices ($m$ for Alice and $n$ for Bob) is
\begin{equation}
J(\mathbf{H})=\frac{2\pi}{mn} \sum_{i=1}^m\sum_{j=1}^n \vert H_{ij} \vert,
\label{eq:J}
\end{equation}
for the B-H system.
One can compare this control cost with the control cost for a single round protocol defined in (\ref{eq:control cost2}) and note
that while the number of control input pairs available to Alice and Bob is the same in both cases, the number of distinct controls used by each is typically not.

In order to enable a concrete comparison between single round protocols and protocols based on completely shared information we will use the B-H system as an example.  First of all, we consider the case that $\mathbf{H}$ is a Hadamard matrix of
order $n$.  Since the entries of such matrices are either 1 or -1, the averaged control cost with perfect information
$J(\mathbf{H})$ is simply $2\pi$.  On the other hand, the singular values of $\mathbf{H}$ are all equal to $\sqrt{n}$.
Thus, by Corollary \ref{cor:1} the optimal single round protocol cost $\hat{C}_{\bF_\bB}(\mathbf{H})$ is
\begin{equation}
\left\{
\begin{array}{lll}
\frac{\pi\sqrt{n}}{2} (n+2) & & {\rm for \,  even \,} n,\\
\frac{\pi\sqrt{n}}{2}(n+2+\frac{1}{n}) &  & {\rm for \, odd \,} n.
\end{array}
\right.
\end{equation}
Thus, the ratio of two control costs is asymptotically $(n+2)\sqrt{n}/4$.

We can perform a similar comparison for the case of orthogonal matrices.   To do so, we need to quote the following bounds on the sum of the absolute
value of the entries in an orthogonal matrix.   These elementary results on matrices are provided here for the sake of completeness.

\begin{proposition}
Consider B-H system.  For an orthogonal $\mathbf{H}$,
\begin{equation}
\frac{2\pi}{n} \leq \frac{2\pi}{n^2} \sum_{i,j=1}^n \vert H_{ij} \vert \leq \frac{2\pi}{\sqrt{n}}.
\end{equation}
For any square matrix $\mathbf{H}$,
\begin{equation}
\frac{2\pi}{n^2} \sum_{i,j=1}^n \vert H_{ij} \vert \leq \frac{2\pi}{n} \sum_{k=1}^n \sigma_k(\mathbf{H}).
\end{equation}
\end{proposition}

\noindent
\textit{Proof}: \rm
To prove the lower bound when $\mathbf{H}$ is an orthogonal matrix, use the inequality
\begin{equation}
\sum_{i,j=1}^n \vert H_{ij} \vert \geq \sum_{i,j=1}^n H_{ij}^2 =n.
\end{equation}
To prove the upper bound, consider the optimization problem,
\begin{equation}
S=\max \sum_{i,j=1}^n x_{ij}
\end{equation}
subject to the constraint that for all $j$,
$\sum_{i=1}^n x_{ij}^2 =1$.
Since the constraint is weaker than the requirement that the $x_{ij}$'s form an orthogonal matrix,
it follows that
\begin{equation}
\sum_{i,j=1}^n \vert H_{ij} \vert \leq S = n \max_{z_1^2+\cdots +z_n^2=1} \sum_{i=1}^n z_i = n\sqrt{n}.
\end{equation}
For the case where $\mathbf{H}$ is a general square matrix, decompose it by SVD so that
\begin{equation}
\mathbf{H}=\mathbf{\Phi}\mathbf{\Lambda}\mathbf{\Theta}, \;\;
\mathbf{\Lambda}=\left[
\begin{array}{cccc}
\sigma_1(\mathbf{H}) & 0 & \cdots & 0 \vspace{-.1cm}\\
0 & \sigma_2(\mathbf{H}) & \cdots & 0 \vspace{-.1cm}\\
\vdots & \vdots & \vdots & \vdots \vspace{-.1cm}\\
0 & 0 & \cdots &  \sigma_n(\mathbf{H})
\end{array} \right].
\end{equation}
where $\mathbf{\Phi}$ and $\mathbf{\Theta}$ are orthogonal.
For $k, 1 \leq k \leq n$, define an $n$-by-$n$ matrix,  $\mathbf{H}(k)$, by
\begin{equation}
\mathbf{H}(k) = \mathbf{\Phi}\mathbf{\Lambda}_k\mathbf{\Theta},\;
\mathbf{\Lambda}_k = diag [ \underbrace{0, \ldots, 0}_{k-1}, \sigma_k(\mathbf{H}),
\underbrace{0, \ldots, 0}_{n-k} ],
\end{equation}
Then,
\begin{equation}
\begin{array}{lll}
\sum_{i,j=1}^n \vert H(k)_{ij} \vert =  \sigma_k(\mathbf{H})
\sum_{i,j=1}^n \vert \Phi_{ik}\Theta_{kj} \vert \\
 \leq  \sigma_k(\mathbf{H}) \sum_{i=1}^n \vert \Phi_{ik} \vert \sum_{i=1}^n \vert \Theta_{kj} \vert
 \leq  n \sigma_k(\mathbf{H}).
\end{array}
\end{equation}
From this, it follows that
\begin{equation}
\frac{2\pi}{n^2} \sum_{i,j=1}^n \vert H_{ij} \vert \leq
\frac{2\pi}{n^2}  \sum_{k=1}^n \sum_{i,j=1}^n \vert H(k)_{ij} \vert
\leq \frac{2\pi}{n} \sum_{k=1}^n \sigma_k(\mathbf{H}).
\end{equation}
\hfill $\blacksquare$

This result together with Corollary \ref{cor:1} imply the following
\begin{theorem}
For the B-H system and a general $n$-by-$n$ target matrix $\mathbf{H}$,
\begin{equation}
\hat{C}_{\bF_\bB}(\mathbf{H})- J(\mathbf{H}) \geq  \frac{2\pi}{n}  \sum_{k=1}^{n} (\lceil k/2 \rceil -1) \sigma_{k}(\mathbf{H}) \geq 0.
\end{equation}
The difference in the control energy is strictly positive if the rank of $\mathbf{H}$ is larger than 2.
\end{theorem}

If $\mathbf{H}$ is an orthogonal matrix, then the optimal single round cost $\hat{C}_F(\mathbf{H})$ is
\begin{equation}
\left\{
\begin{array}{lll}
\frac{\pi}{2} (n+2) & & {\rm for \, even \,} n,\\
\frac{\pi}{2}(n+2+\frac{1}{n}) & & {\rm for \, odd \,} n.
\end{array}
\right.
\end{equation}
By comparison,
\begin{equation}
\frac{2\pi}{n} \leq J(\mathbf{H}) \leq \frac{2\pi}{\sqrt{n}}.
\end{equation}

For example, consider the case where $n=2$.  The
identity matrix, $\mathbf{I}_2$, incurs an averaged cost of $\pi$ under information sharing:
there are two entries whose computation requires control energy of $2\pi$; for the other
two entries, $H_{12}$ and $H_{21}$, zero control can be used.
However, for the case where the information on the choice is not shared,
it is not possible to save control cost by setting any of the controls to zero, resulting in a cost increase by a factor of 2.
In general, the control cost ratio grows super-linearly as a function of the dimension of target matrix.

\section{MULTI-ROUND PROTOCOLS}\setcounter{equation}{0}

Analysis from the previous section indicates that control cost can be substantially reduced by using protocols that allow communication between the agents.   For the extreme scenario where the agents have complete prior information of
each others inputs, optimization of the control cost can be reduced to solving a family of single output target optimization
problems.  For any $m$-by-$n$ target matrix $\bf H$ and a general bilinear
input-output mapping, $\bf F$, the averaged control cost for solving a family of single
output target optimization problems is given by:
\begin{equation}
J(\bH)=\frac{2}{mn\sigma_{1}(\mathbf{F})} \sum_{i=1}^{m} \sum_{j=1}^{n} |H_{ij}|.
\label{eq:cost}
\end{equation}
This is a generalization of equation (\ref{eq:J}).

If the agents do not have prior information on the inputs to be selected, they can
communicate their input choice to each other in a multi-round protocol.
Detailed analysis of general multi-round protocols lies beyond the scope of this paper.  We shall briefly consider two-phase protocols, however, in which one phase allows partial information to be shared at negligible cost.  The main result here is that if the cost of signaling certain partial information is negligible, the control cost can be made arbitrarily
close to the lower bound, $J(\bH)$ that was determined in the previous section.   It is important to note that the protocols approaching $J(\bH)$ do not necessarily
require the agents to communicate their choices completely to each other.  It will be shown that the number of bits communicated is related to the classical communication complexity of computing $\bf H$.  This result provides insight into the {\em value of information} in terms of the control energy savings that can be achieved through communication between the agents as they evaluate $\bf H$.

Various concepts of communication {\em at negligible cost} can be considered.  One possible approach to partial information exchange in a two-phase protocol is to use a {\em side channel} in an initiation phase in which it is assumed that there is no cost of transmitting some partial information between the agents.  It is in the second phase that  this information is used by the agents to select controls that effect the computation specified in the single-round protocols of Section 5.  Another approach to information sharing at negligible communication cost is to assume that under certain circumstances, very low cost control signals can be used.  Formally, 
for the model (\ref{eq:basic1}) described in section 2, we introduce the following concept of
{\em $\epsilon$-signaling capability}.

\begin{definition}
Alice possesses $\epsilon$-signaling capability around the initial state $\bx_0$ if for any $\epsilon$ there exist times, $0 < t_1 < t_2$ and controls
$u_1$ and $u_2$ for Alice, $v_1$ for Bob, so that
\begin{equation}
\int_0^{t_2} (u_1^2+v_1^2)dt < \epsilon, \; \int_0^{t_2} (u_2^2+v_1^2)dt < \epsilon.
\end{equation}
Moreover, when $u$ is set to $u_1$ or $u_2$ and $v$ is set to $v_1$, the following conditions hold:

\noindent
{\rm 1)} $\bx_{u_1,v_1}(t_1) \neq \bx_{u_2,v_1}(t_1)$,
{\rm 2)} $\bx_{u_1,v_1}(t_2)=\bx_{u_2,v_1}(t_2)=\bx_0$.

\end{definition}
One can define similar capability for Bob.  For example, for the B-H system both agents
possess $\epsilon$-signaling capability as the loop controls can enclose arbitrarily small areas.
For systems in which both agents have $\epsilon$-signaling capability one can design
multi-round protocols for realizing $\bH$ with control cost arbitrarily close to $J(\bH)$.
These protocols consist of two phases.
In the first phase, the agents communicate their choices of inputs to each other.
Based on the information received, the original target matrix is decomposed into a finite number
of sub-matrices and controls can then be applied to realize the sub-matrix that corresponds to the
choices of the agents.  We call such a protocol a {\em two-phase protocol.}
To describe the detail, we recall the definition of a {\em monochromatic matrix}, (see for example \cite{KN}).

\begin{definition}
A sub-matrix is said to be monochromatic if all its entries have the same value.
\end{definition}

Given an $m$-by-$n$ target matrix, $\mathbf{H}$, we can define a set of sub-matrices, $\{ \mathbf{H}_1, \ldots, \mathbf{H}_K \}$,
so that $ \mathbf{H}_k$ is an $m_k$-by-$n_k$ sub-matrix with its $(i,j)$ entry defined by
\begin{equation}
H_k (i,j)= H_{t_{k,i}, s_{k,j}}
\end{equation}
where $t_{k,i}$ lies in an index set $\mathcal{M}_k \subset \{1, \ldots, m \}$
and $s_{k,j}$ lies in an index set $\mathcal{N}_k \subset \{1, \ldots, n \}$.   Note that by definition,
$|\mathcal{M}_k|=m_k,  |\mathcal{N}_k| =n_k$.   Define $l_k = min (m_k, n_k)$.

The set of sub-matrices forms a matrix partition for $\mathbf{H}$ if the following holds:
\begin{enumerate}
\item
For any $i$ and $j$ with $1\leqslant i \leqslant m$ and $ 1\leqslant j \leqslant n$, there exist $k$, $\alpha$, and $\beta$ such that
$t_{k,\alpha}=i, \; s_{k,\beta}=j$.
\item If $t_{k,\alpha}=t_{k',\alpha'}, \; s_{k,\beta}=s_{k',\beta'}$, then
$k=k'$,  $\alpha=\alpha'$, and $\beta=\beta'$.
\end{enumerate}

For example, the following figure shows a sub-matrix partition involving five sub-matrices:
\begin{equation}
\left[ 
\begin{array}{cc}
\begin{array}{c}
\fbox{$\begin{array}{cc}
        \vspace{.8mm} H_{11} \:& \:H_{12} \vspace{1.2mm}\\
        H_{21} \:& \: H_{22}
			\end{array}$} \vspace{.5mm}\\
\fbox{$\begin{array}{c}
        H_{31} \vspace{.5mm}\\
        H_{41}
			\end{array}$}\,
\fbox{$\begin{array}{c}
        H_{31} \\
        H_{41}
			\end{array}$}
\end{array}
& \hspace{-.5mm}
\begin{array}{cc}
\fbox{$\begin{array}{cc}
			H_{13} & H_{14}
			\end{array}$} & \vspace{.7mm}\\
\fbox{$\begin{array}{cc}
        H_{23} & H_{24}\vspace{1.5mm}\\ 
			H_{33} & H_{34}\vspace{1.5mm}\\
			H_{43} & H_{44}\\ 
			\end{array}$}&
\end{array}
\end{array}
\right]
\label{eq:partition}
\end{equation}

It follows from direct verification that for all $k$
\begin{align}
\sum_{l=1}^{l_k} \sigma_l^2 (\mathbf{H}_k) &= tr \mathbf{H}_k\mathbf{H}_k^T
=\sum_{i=1}^{m_k} \sum_{j=1}^{n_k} H_k^2 (i,j)\\
&=\sum_{t_{k,i} \in \mathcal{M}_k} \sum_{s_{k,j} \in \mathcal{N}_k} H_{t_{k,i},s_{k,j}}^2.
\label{eqn:prop6.2}
\end{align}
Thus, for a sub-matrix partition into $K$ sub-matrices,
\begin{equation}
\sum_{k=1}^K \sum_{l=1}^{l_k} \sigma_l^2(\mathbf{H}_k) = \sum_{i=1}^{m} \sum_{j=1}^{n} H^2_{i,j} = tr \mathbf{H}\mathbf{H}^T.
\label{eqn:prop6.3}
\end{equation}

A sub-matrix partition can be regarded as a decomposition of a complex distributed control problem into simpler
sub-problems.  We can estimate the effectiveness of a decomposition by calculating the control cost averaged
over the decomposed sub-problems, namely,
\begin{equation}
A  = \frac{1}{mn} \sum_{k=1}^K m_k n_k \hat{C}_{\mathbf{F}}(\bH_k).
\end{equation}
The following result provides a lower bound for this averaged control cost.

\begin{theorem}
\label{thm:2}
Consider a bounded, bilinear input-output mapping, $F$, with a regular matrix representation $\mathbf{F}$.
The average control cost, $A$, for applying single round protocols to the sub-matrices in a sub-matrix partition
$\{ \mathbf{H}_1, \ldots, \mathbf{H}_K \}$ of $\mathbf{H}$ satisfies the lower bound:
\begin{equation}
A  = \frac{1}{mn} \sum_{k=1}^K m_k n_k \hat{C}_{\mathbf{F}}(\bH_k)
\geq 
\frac{2}{mn\sigma_{1}(\mathbf{F})} \sum_{i=1}^{m} \sum_{j=1}^{n} |H_{ij}|.
\label{eq:thm7.1}
\end{equation}
If all the sub-matrices are monochromatic, this lower bound is the infimum value of $A$.
\end{theorem}

\noindent
\textit{Proof}: \rm
\begin{align}
A& =  \frac{1}{mn} \sum_{k=1}^K m_k n_k \hat{C}_{\mathbf{F}}(\bH_k)\\
& \geq  \frac{2}{mn} \sum_{k=1}^K \sqrt{m_k n_k}
\sum_{l=1}^{l_k} \sigma_{l}(\mathbf{H}_k)/\sigma_{l}(\mathbf{F})   \label{ieq:1}\\
&\geq \frac{2}{mn\sigma_{1}(\mathbf{F})} \sum_{k=1}^K \sqrt{m_k n_k} \sum_{l=1}^{l_k} \sigma_{l}(\mathbf{H}_k)
\label{ieq:2}\\
& \geq \frac{2}{mn\sigma_{1}(\mathbf{F})} \sum_{k=1}^K \left( m_k n_k
\sum_{l=1}^{l_k} \sigma_{l}^2(\mathbf{H}_k) \right)^{1/2}\label{ieq:3}\\
&=\frac{2}{mn\sigma_{1}(\mathbf{F})} \sum_{k=1}^K \left( m_k n_k
 \sum_{i=1}^{m_k} \sum_{j=1}^{n_k} H_k^2 (i,j) \right)^{1/2}  \\
& \geq \frac{2}{mn\sigma_{1}(\mathbf{F})} \sum_{k=1}^K
\sum_{i=1}^{m_k} \sum_{j=1}^{n_k} |H_k (i,j)| \label{ieq:4}\\
& = \frac{2}{mn\sigma_{1}(\mathbf{F})} \sum_{i=1}^{m} \sum_{j=1}^{n} 
|H_{ij}| \label{ieq:5}.
\end{align}

(\ref{ieq:4}) follows from the well-known inequality that for any real numbers, $(x_1, \ldots, x_p)$
\begin{equation}
p\sum_{i=1}^p x_i^2 \geq \left( \sum_{i=1}^p x_i \right)^2
\label{ieq:6}
\end{equation}
with equality holding if and only if all the $x_i$'s equal to each other.

According to Theorem \ref{thm:1} the infimum value of $A$ is given by the right-hand-side of (\ref{ieq:1}).
If all the sub-matrices are monochromatic, then
$ \sigma_{l}(\mathbf{H}_k)$ if $l >1$.  Hence inequalities (\ref{ieq:2}), (\ref{ieq:3}), and (\ref{ieq:4}) all become equality, and
the last expression in (\ref{ieq:5}) is the infimum value.
\hfill $\blacksquare$

In the first phase of a two-phase protocol, the agents communicate with each other via the dynamic system
by means of $\epsilon$ signals.  The bit sequence defined by the communication complexity
protocol can be regarded as an algorithm to identify the chosen sub-matrix in a given
partition.  We can visualize the algorithm by moving down a binary tree,
so that depending on the value of the bit sent by either one of the agents in the
communication protocol, we descend from a given node to its left or right child.   For additional details about communication protocol,we refer to \cite{KN} or \cite{Wong}.  To make explicit contact with \cite{KN}, suppose that each sub-matrix in the given matrix partition is monochromatic.  The number of leaves in the binary tree that defines the protocol is equal to the number of sub-matrices defining the partition, and each of these sub-matrices is mapped to one of the leaves of the binary tree.  The maximum number of bits communicated in the protocol is equal to two times the depth of the tree.   (Since the communication has to pass through the dynamic system, if Alice wants to send one bit of information to Bob, the bit has to pass from Alice to the dynamic system and from the dynamic system to Bob, leading to two communication bits being exchanged.)  The {\em protocol complexity} is thus defined as two times the height of the binary tree and can provide an upper bound for control communication complexity. 

For illustration, consider a target function,
\begin{equation}
\bf{H} = \left[
\begin{array}{cccc}
1 & 1 & 1 & 1 \vspace{-.2cm}\\
1 & 1 & 5 & 5 \vspace{-.2cm}\\
2 & 3 & 5 & 5 \vspace{-.2cm}\\
2 & 3 & 5 & 5 \\ 
\end{array}\right]
\label{eq:jb:target}
\end{equation}
The sub-matrix partition shown in (\ref{eq:partition}) represents a decomposition of the target matrix
into the minimum number of monochromatic blocks.  (Such minimal decomposition is not unique.)
One can define a communication protocol to identify the components of the matrix partition (monochromatic blocks in this case) as follows.  

\bigskip
\noindent{\bf Protocol to communicate the structure of a matrix partition:} Assume that Alice controls the choice of columns and Bob controls
the choice of rows.

\begin{enumerate}
\item Alice sends a bit to Bob with value 0 if she chooses the first 2 columns, otherwise she sends 1.
\item Upon receiving a bit of value 1 from Alice, Bob sends a bit to Alice with value 0 if he chooses
the first row, otherwise, he sends a bit of value 1.  No further bit needs to be sent.
\item On the other hand if a bit of value 0 is received from Alice, Bob sends a bit to Alice with
value 0 if he chooses the first two rows, otherwise he sends a bit 1.  Only in the latter case, Alice sends
one more bit, with value 0 if she chooses the first column and 1 if she chooses the second column.
\end{enumerate}

\begin{figure}[h] 
\begin{center}
\vspace{-.1cm}
\includegraphics[width=4.5in]{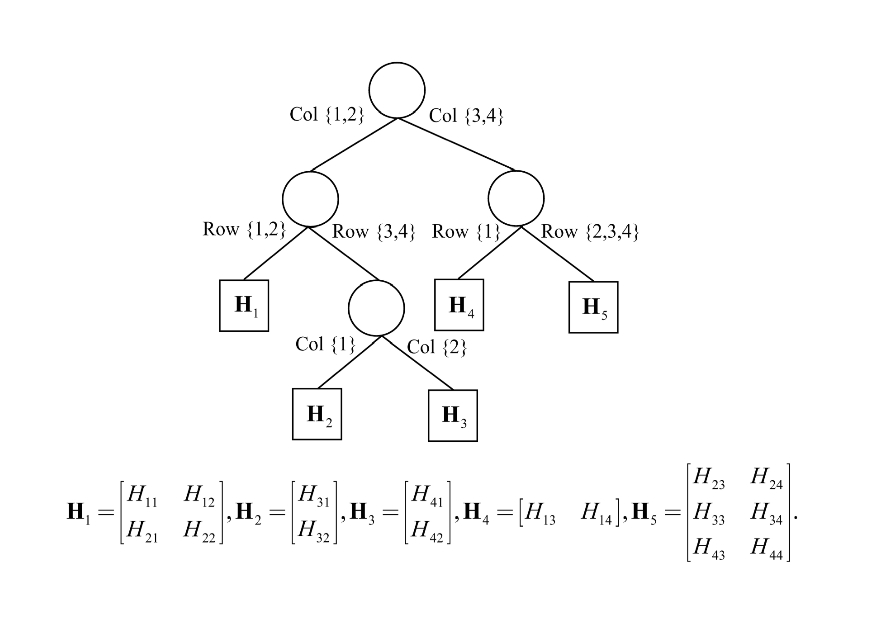} 
\vspace{-0.8cm}
\caption{Binary tree of a communication protocol to realize the partition
defined by the matrix in equation (\ref{eq:partition}).}  \label{fig1}
\end{center}
\end{figure}

The communication protocol can be represented by the binary tree shown in Figure \ref{fig1}.
A maximum of six bits (counting bits sent by the dynamic system) are needed in this protocol
in order to guarantee all sub-matrices can be identified. 

In our two-phase protocol, once the communication phase is completed---which is to say that the phase-one protocol has run to completion and a leaf node identifying a sub-matrix has been reached, the second phase of the protocol starts.  The target matrix that is collaboratively evaluated by the agents in phase two is the sub-matrix that had been selected in phase one.  It is assumed that in phase one, communication of negligible cost (e.g.\ $\epsilon$-signaling) occurs, but that in phase two, open-loop controls of the form described in Section 5 realize the output specified in the chosen sub-matrix.
It is clear that one can construct two-phase protocols with total control cost arbitrarily
close to (\ref{eq:cost}), the lower bound for all protocols realizing $\bH$.		
We summarize the results in this section in the following theorem.

\begin{theorem}
\label{thm:3}
Consider a bounded, bilinear input-output mapping, $F$, with a regular matrix representation $\mathbf{F}$.
Suppose both agents have $\epsilon$-signaling capability around the initial state.
Let $\mathbf{H}$ be an $m$-by-$n$ target matrix.
The infimum control cost of any multi-round protocol that realizes $\mathbf{H}$ is given by:
\begin{equation}
J(\bH)=\frac{2}{mn\sigma_{1}(\mathbf{F})} \sum_{i=1}^{m} \sum_{j=1}^{n} |H_{ij}|.
\label{eq:thm7.2}
\end{equation}
\end{theorem}

By comparing the control cost of an optimal single round protocol as given by (\ref{eqn:thm1}) with that of the multi-round protocol (\ref{eq:thm7.2}), one
can estimate the value of  communicated bits in reducing the control energy cost.  
Using the target matrix $\bf{H}$ of (\ref{eq:jb:target}) as an example, for the B-H system the minimal control cost without using any communication is defined in equation (\ref{eq:thm5.1}) and has a value of  $7.68 \pi$.  If
communication cost is negligible then the minimal control cost is approximated by the right-hand-side of (\ref{eq:thm7.1}) and has a value of $2.88 \pi$.  This can be achieved by the information sharing protocol shown in Figure \ref{fig1}, which has a protocol complexity of 6 ($=$ two times the length of the binary tree).  The protocol complexity is a ``worst case'' metric, reflecting the maximum number of bits that might need to be communicated.  Thus, the value of a single communication bit in reducing the control energy cost for this problem is at least $0.8 \pi$ with the units being control energy (as defined by (\ref{eq:control cost})) per bit.

\section{CONCLUSION}\setcounter{equation}{0}

This paper has continued our study of problems in control communication complexity, which may be viewed as an extension of classical communication complexity with the additional focus on control cost.  There are several important application contexts in which the optimization problems of the type we have considered seem to arise naturally.  The single round protocols for steering the B-H system realizes the solution to a problem in distributed computing where independent agents act to evaluate a function without foreknowledge of each other's choices.  As was noted in \cite{CDC09}, the problems also arise naturally in what we have called the {\em standard parts optimal control problem} in which it is desired to find a specific number, $m$, of control inputs to a given input-output system that can be used in different combinations to attain a certain number, $n$, of output objectives so as to minimize the cost of control averaged over the different objectives. 
The connection with quantum control, and in particular, the control of quantum spin systems (see for example the references to quantum control and computation in \cite{WB}) is another interesting line of investigation that is under way and will be treated elsewhere.

\appendices
\section{}\setcounter{equation}{0}
\setcounter{section}{8}

To prove the rank of a matrix representation of a bilinear input-output mapping is well-defined, let
$\mathbf{F}$ be a representation corresponding to the bases $\{e_1, e_2, \ldots \}$
and $\{f_1, f_2, \ldots \}$.  Let $\mathbf{F}'$ be a representation corresponding to the bases $\{ e_1, e_2, \ldots \}$
and $\{ f'_1, f'_2, \ldots \}$.  Assume first that $\mathbf{F}$ has finite rank equal to $r$.
By relabeling the indices, we can assume without loss of generality that the first $r$ columns of $\mathbf{F}$
are independent.  Hence, there exists a matrix $\mathbf{C}$ with $r$ rows and infinitely many columns such that
\begin{align}
\mathbf{F} &=
\left[
\small
\begin{array}{llll}
F_{11} & F_{12} & F_{13} &  \ldots \\
F_{21} & F_{22}  & F_{23} &  \ldots\\
F_{31} & F_{32} & F_{33} &  \ldots \\
\vdots & \vdots & \vdots & \ddots\\
\end{array}
\right]\\
&= \left[
\begin{array}{llll}
F_{11} & F_{12} &  \ldots & F_{1r}\\
F_{21} & F_{22} &  \ldots & F_{2r}\\
F_{31} & F_{32} &  \ldots & F_{3r}\\
\vdots & \vdots & \vdots & \vdots\\
\end{array}
\right] \mathbf{C} \equiv \mathbf{\bar{F}}\mathbf{C} .
\end{align}
By relabeling the indices of the basis $\{e_1, e_2, \ldots \}$ if necessary, we can assume that the
first $r$ rows $\mathbf{\bar{F}}$ are independent.  Label the sub-matrix of $\mathbf{\bar{F}}$ formed by
the first $r$ rows by $\mathbf{D}$.  Note that $\mathbf{D}$ is an invertible $r$-by-$r$ matrix and
\begin{equation}
\mathbf{C}=
\mathbf{D}^{-1} \left[
\small
\begin{array}{llll}
F_{11} & F_{12} & F_{13} &  \ldots \\
F_{21} & F_{22}  & F_{23} &  \ldots\\
\vdots & \vdots & \vdots & \vdots\\
F_{r1} & F_{r2} & F_{r3} &  \ldots \\
\end{array}
\right].
\end{equation}
That is, for $1 \leq k \leq r$ and $1 \leq l$
\begin{equation}
C_{kl}=\sum_{i=1}^r D^{-1}_{ki}F_{il}.
\label{eq:C}
\end{equation}
Let $C_0$ be the upper bound of the entries in $\mathbf{D}^{-1}$, so that for all $1 \leq i,j \leq r$,
$D_{ij}^{-1} \leq C_0 < \infty.$
For any positive integer $j$, the base vector, $f'_j$, has a representation of the form
\begin{equation}
f'_j = \sum_{l=1}^\infty \alpha_{jl}f_l.
\end{equation}
The coefficients are square summable so that
$\sum_{l=1}^\infty \alpha_{jl}^2 < \infty.$
For any $1 \leq i \leq r$, define
\begin{equation}
g_{ij} = \sum_{l=1}^\infty \alpha_{jl}sgn(\alpha_{jl}F_{il})f_l.
\end{equation}
Since the coefficients for $g_{ij}$ are square summable, this is a well-defined element in $\mathcal{L}$.
Moreover,
\begin{equation}
F(e_i, g_{ij}) = \sum_{l=1}^\infty \vert \alpha_{jl}F_{il} \vert < \infty,
\end{equation}
since $F$  is bounded.  On the other hand,
\begin{equation}
F'_{ij}=F(e_i, f'_j) = \sum_{l=1}^\infty \alpha_{jl}F_{il}
=\sum_{l=1}^\infty \alpha_{jl} \sum_{k=1}^r F_{ik}C_{kl}.
\label{eq:F1}
\end{equation}
Formally, by exchange the summation order, the right-hand-side of equation (\ref{eq:F1}) is equal to
\begin{equation}
\sum_{k=1}^r F_{ik} \sum_{l=1}^\infty \alpha_{jl}C_{kl}
= \sum_{k=1}^r F_{ik}E_{jk}, \; \;  E_{jk} \equiv \sum_{l=1}^\infty \alpha_{jl}C_{kl}.
\label{eq:F2}
\end{equation}
The equality holds if the expression in (\ref{eq:F2}) is convergent.
By means of equation (\ref{eq:C}),
\begin{equation}
\vert E_{jk} \vert  \leq
\sum_{l=1}^\infty \sum_{i=1}^r \vert \alpha_{jl}D^{-1}_{ki}F_{il} \vert
\leq    C_0 \sum_{i=1}^r \sum_{l=1}^\infty   \vert \alpha_{jl}F_{il} \vert.
\end{equation}
It follows that the expression in (\ref{eq:F2}) is convergent and
\begin{equation}
F'_{ij}
=\sum_{l=1}^\infty \alpha_{jl} \sum_{k=1}^r F_{ik}C_{kl}
=\sum_{k=1}^r F_{ik} \sum_{l=1}^\infty \alpha_{jl}C_{kl}.
\label{eq:F3}
\end{equation}
Then equation (\ref{eq:F3}) can be rewritten as
$\mathbf{F}'=\mathbf{\bar{F}}\mathbf{E}, \;
\mathbf{E} \equiv
\left[
E_{ij}
\right].$
Therefore, the rank of $\mathbf{F}'$ is equal to or less than $r$.  By reversing the roles of $\mathbf{F}$ and $\mathbf{F}'$,
the ranks of the two matrices must be identical if one of them has finite rank.   Thus, if
the basis $\{e_1, e_2, \ldots \}$ is fixed the rank is independent of the other basis. 
Similarly, one can argue that if $\{f_1, f_2, \ldots \}$ is fixed, the rank is independent of the other basis.

In the general case, consider matrix representation, $\mathbf{F}$, with respect to bases
$\{e_1, e_2, \ldots \}$ and $\{f_1, f_2, \ldots \}$, and  matrix representation, $\mathbf{F}''$, with respect to bases
$\{e'_1, e'_2, \ldots \}$ and $\{f'_1, f'_2, \ldots \}$.  If the rank of $\mathbf{F}$ is finite, it is equal to the rank of the
matrix representation with respect to the bases $\{e_1, e_2, \ldots \}$  and $\{f'_1, f'_2, \ldots \}$, which in turn
is equal to the rank $\mathbf{F}''$.  It is easy to see that it is not possible for one representation to have finite
rank while the other has an infinite rank.  Thus the rank is independent of the choice of the bases.

\setcounter{section}{1}
\section{}\setcounter{equation}{0}
\setcounter{section}{9}

Let $\{e_1, e_2, \ldots \}$ and $\{f_1, f_2, \ldots \}$ be orthonormal bases corresponding to the representation $F$.
Let  $v=\sum_{j=1}^l v_j f_j$ be an element of $\mathcal{L}$ and
${\bf v}^T = [v_1, \ldots , v_l]$ be the corresponding $l$-dimensional vector.
Let ${\bf u}^T = {\bf v}^T {\bf F}_l^T$
and denote the corresponding element $\sum_{i=1}^l u_i e_i$
in $\mathcal{L}$ by $u$.  Then,
\begin{equation}
F(u,v) = {\bf u}^T {\bf F}_l {\bf v} = {\bf v}^T {\bf F}_l^T {\bf F}_l {\bf v}.
\end{equation}
Using the assumption that F is bounded and the inequality (\ref{eq:boundedF}), one can
show that
\begin{align}
\| {\bf F}_l {\bf v} \|^2 &= {\bf v}^T {\bf F}_l^T {\bf F}_l {\bf v}  = F(u,v)
\leq \|F\|  \|{\bf v}\| \|{\bf F}_l {\bf v}\|\\
&\leq \|F\|^2 \|{\bf v}\|^2.
\end{align}

Let  ${\bf w}_i$ be the eigenvector of ${\bf F}_l^T {\bf F}_l$ with unit norm
that corresponds to $\sigma_i({\bf F}_l)$.  Then,
\begin{equation}
\sigma_i^2({\bf F}_l) =  {\bf w}_i^T {\bf F}_l^T {\bf F}_l {\bf w}_i  = \| {\bf F}_l {\bf w}_i \|^2
\leq  \| F \|^2.
\end{equation}
\vspace{-0.5in}


\end{document}